\documentclass[twocolumn]{aastex631}

\accepted{April 1, 2024}

\shorttitle{MAHA IV}
\shortauthors{Zastrocky et al.}

\begin{document}

\title{Monitoring AGNs with H$\beta$ Asymmetry. IV. First Reverberation Mapping Results of 14 AGNs}

\correspondingauthor{T.E. Zastrocky}
\email{tzastroc@uwyo.edu}

\author{T.E. Zastrocky}
\affiliation{Department of Physics and Astronomy, University of Wyoming, Laramie, WY 82071, USA} 

\author{Michael S. Brotherton}
\affiliation{Department of Physics and Astronomy, University of Wyoming, Laramie, WY 82071, USA} 

\author{Pu Du}
\affiliation{Key Laboratory for Particle Astrophysics, Institute of High Energy Physics, Chinese Academy of Sciences, 19B Yuquan Road, Beijing 100049, People's Republic of China}

\author{Jacob N. McLane}
\affiliation{Department of Physics and Astronomy, University of Wyoming, Laramie, WY 82071, USA} 

\author{Kianna A. Olson}
\affiliation{Department of Physics and Astronomy, University of Wyoming, Laramie, WY 82071, USA} 

\author{D.A. Dale}
\affiliation{Department of Physics and Astronomy, University of Wyoming, Laramie, WY 82071, USA} 

\author{H.A. Kobulnicky}
\affiliation{Department of Physics and Astronomy, University of Wyoming, Laramie, WY 82071, USA} 

\author{Jaya Maithil}
\affiliation{Center for Astrophysics $|$ Harvard \& Smithsonian, 60 Cambridge Street, Cambridge, MA 02138, USA.} 

\author{My L. Nguyen}
\affiliation{Department of Physics and Astronomy, University of Wyoming, Laramie, WY 82071, USA} 

\author{William T. Chick}
\affiliation{Department of Physics and Astronomy, University of Wyoming, Laramie, WY 82071, USA} 

\author{David H. Kasper}
\affiliation{Department of Physics and Astronomy, University of Wyoming, Laramie, WY 82071, USA} 

\author{Derek Hand}
\affiliation{Department of Physics and Astronomy, University of Wyoming, Laramie, WY 82071, USA} 

\author{C. Adelman}
\affiliation{Department of Physics and Astronomy, University of Wyoming, Laramie, WY 82071, USA} 
\affiliation{Department of Physics and Astronomy, Cal Poly Pomona, Pomona, CA 91768, USA}

\author{Z. Carter}
\affiliation{Department of Physics and Astronomy, University of Wyoming, Laramie, WY 82071, USA} 
\affiliation{Department of Physics and Astronomy, Trinity University, San Antonio, TX 78212, USA}

\author{G. Murphree}
\affiliation{Department of Physics and Astronomy, University of Wyoming, Laramie, WY 82071, USA}
\affiliation{Institute for Astronomy, University of Hawai’i, Honolulu, HI 96822, USA}

\author{M. Oeur}
\affiliation{Department of Physics and Astronomy, University of Wyoming, Laramie, WY 82071, USA} 
\affiliation{Department of Physics and Astronomy, California State University, Long Beach, Long Beach, CA 90840, USA}

\author{T. Roth}
\affiliation{Department of Physics and Astronomy, University of Wyoming, Laramie, WY 82071, USA} 
\affiliation{Department of Physics and Astronomy, California State University, Sacramento, CA 95747, USA}

\author{S. Schonsberg}
\affiliation{Department of Physics and Astronomy, University of Wyoming, Laramie, WY 82071, USA}
\affiliation{Department of Physics and Astronomy, University of Montana, Missoula, MT 59812, USA}

\author{M.J. Caradonna}
\affiliation{Department of Physics and Astronomy, University of Wyoming, Laramie, WY 82071, USA} 
\affiliation{Department of Physics, University of North Texas, Denton, Texas 76203, USA}

\author{J. Favro}
\affiliation{Department of Physics and Astronomy, University of Wyoming, Laramie, WY 82071, USA} 
\affiliation{Department of Astronomy, Diablo Valley College, Pleasant Hill, CA 94523, USA} 

\author{A.J. Ferguson}
\affiliation{Department of Physics and Astronomy, University of Wyoming, Laramie, WY 82071, USA} 

\author{I.M. Gonzalez}
\affiliation{Department of Physics and Astronomy, University of Wyoming, Laramie, WY 82071, USA} 
\affiliation{Department of Physics, Oklahoma Baptist University, Shawnee, OK 74804, USA}

\author{L.M. Hadding}
\affiliation{Department of Physics and Astronomy, University of Wyoming, Laramie, WY 82071, USA}
\affiliation{Department of Physics, Grinnell College, Grinnell, IA 50112, USA}

\author{H.D. Hagler}
\affiliation{Department of Physics and Astronomy, University of Wyoming, Laramie, WY 82071, USA} 
\affiliation{Department of Astronomy, Whitman College, Walla Walla, WA 99362, USA}

\author{C.J. Rogers}
\affiliation{Department of Physics and Astronomy, University of Wyoming, Laramie, WY 82071, USA} 
\affiliation{Department of Physics and Astronomy, Cal Poly Humboldt, Arcata, CA 95521, USA}

\author{T.R. Stack}
\affiliation{Department of Physics and Astronomy, University of Wyoming, Laramie, WY 82071, USA}
\affiliation{Department of Physics, Illinois Institute of Technology, Chicago, IL 60616, USA}

\author{Franklin Chapman}
\affiliation{Department of Physics and Astronomy, University of Wyoming, Laramie, WY 82071, USA}

\author{Dong-Wei Bao}
\affiliation{Key Laboratory for Particle Astrophysics, Institute of High Energy Physics, Chinese Academy of Sciences, 19B Yuquan Road, Beijing 100049, People's Republic of China} 
\affiliation{School of Astronomy and Space Science, University of Chinese Academy of Sciences, 19A Yuquan Road, Beijing 100049, People's Republic of China}

\author{Feng-Na Fang}
\affiliation{Key Laboratory for Particle Astrophysics, Institute of High Energy Physics, Chinese Academy of Sciences, 19B Yuquan Road, Beijing 100049, People's Republic of China} 
\affiliation{School of Astronomy and Space Science, University of Chinese Academy of Sciences, 19A Yuquan Road, Beijing 100049, People's Republic of China}

\author{Shuo Zhai}
\affiliation{Key Laboratory for Particle Astrophysics, Institute of High Energy Physics, Chinese Academy of Sciences, 19B Yuquan Road, Beijing 100049, People's Republic of China} 
\affiliation{School of Astronomy and Space Science, University of Chinese Academy of Sciences, 19A Yuquan Road, Beijing 100049, People's Republic of China}

\author{Sen Yang}
\affiliation{Key Laboratory for Particle Astrophysics, Institute of High Energy Physics, Chinese Academy of Sciences, 19B Yuquan Road, Beijing 100049, People's Republic of China} 
\affiliation{School of Astronomy and Space Science, University of Chinese Academy of Sciences, 19A Yuquan Road, Beijing 100049, People's Republic of China}

\author{Yong-Jie Chen}
\affiliation{Key Laboratory for Particle Astrophysics, Institute of High Energy Physics, Chinese Academy of Sciences, 19B Yuquan Road, Beijing 100049, People's Republic of China} 
\affiliation{School of Astronomy and Space Science, University of Chinese Academy of Sciences, 19A Yuquan Road, Beijing 100049, People's Republic of China}

\author{Hua-Rui Bai}
\affiliation{Key Laboratory for Particle Astrophysics, Institute of High Energy Physics, Chinese Academy of Sciences, 19B Yuquan Road, Beijing 100049, People's Republic of China} 
\affiliation{School of Astronomy and Space Science, University of Chinese Academy of Sciences, 19A Yuquan Road, Beijing 100049, People's Republic of China}

\author{Yi-Xin Fu}
\affiliation{Key Laboratory for Particle Astrophysics, Institute of High Energy Physics, Chinese Academy of Sciences, 19B Yuquan Road, Beijing 100049, People's Republic of China} 
\affiliation{School of Astronomy and Space Science, University of Chinese Academy of Sciences, 19A Yuquan Road, Beijing 100049, People's Republic of China}

\author{Jun-Rong Liu}
\affiliation{Key Laboratory for Particle Astrophysics, Institute of High Energy Physics, Chinese Academy of Sciences, 19B Yuquan Road, Beijing 100049, People's Republic of China} 
\affiliation{School of Astronomy and Space Science, University of Chinese Academy of Sciences, 19A Yuquan Road, Beijing 100049, People's Republic of China}

\author{Zhu-Heng Yao}
\affiliation{Key Laboratory for Particle Astrophysics, Institute of High Energy Physics, Chinese Academy of Sciences, 19B Yuquan Road, Beijing 100049, People's Republic of China} 
\affiliation{School of Astronomy and Space Science, University of Chinese Academy of Sciences, 19A Yuquan Road, Beijing 100049, People's Republic of China}

\author{Yue-Chang Peng}
\affiliation{Key Laboratory for Particle Astrophysics, Institute of High Energy Physics, Chinese Academy of Sciences, 19B Yuquan Road, Beijing 100049, People's Republic of China} 
\affiliation{School of Astronomy and Space Science, University of Chinese Academy of Sciences, 19A Yuquan Road, Beijing 100049, People's Republic of China}

\author{Yu-Yang Songsheng}
\affiliation{Key Laboratory for Particle Astrophysics, Institute of High Energy Physics, Chinese Academy of Sciences, 19B Yuquan Road, Beijing 100049, People's Republic of China} 
\affiliation{School of Astronomy and Space Science, University of Chinese Academy of Sciences, 19A Yuquan Road, Beijing 100049, People's Republic of China}

\author{Yan-Rong Li}
\affiliation{Key Laboratory for Particle Astrophysics, Institute of High Energy Physics, Chinese Academy of Sciences, 19B Yuquan Road, Beijing 100049, People's Republic of China} 

\author{Jin-Ming Bai}
\affiliation{Yunnan Observatories, Chinese Academy of Sciences, Kunming 650011, People's Republic of China}

\author{Chen Hu}
\affiliation{Key Laboratory for Particle Astrophysics, Institute of High Energy Physics, Chinese Academy of Sciences, 19B Yuquan Road, Beijing 100049, People's Republic of China} 

\author{Ming Xiao}
\affiliation{Key Laboratory for Particle Astrophysics, Institute of High Energy Physics, Chinese Academy of Sciences, 19B Yuquan Road, Beijing 100049, People's Republic of China} 

\author{Luis C. Ho}
\affiliation{Kavli Institute for Astronomy and Astrophysics, Peking University, Beijing 100871, People's Republic of China}
\affiliation{Department of Astronomy, School of Physics, Peking University, Beijing 100871, People's Republic of China}

\author{Jian-Min Wang}
\altaffiliation{PI of the MAHA Project.}
\affiliation{Key Laboratory for Particle Astrophysics, Institute of High Energy Physics, Chinese Academy of Sciences, 19B Yuquan Road, Beijing 100049, People's Republic of China}
\affiliation{School of Astronomy and Space Science, University of Chinese Academy of Sciences, 19A Yuquan Road, Beijing 100049, People's Republic of China}
\affiliation{National Astronomical Observatories of China, Chinese Academy of Sciences, 20A Datun Road, Beijing 100020, People's Republic of China}

\begin{abstract}
We report first-time reverberation mapping results for 14 AGNs from the ongoing Monitoring AGNs with H$\beta$ Asymmetry campaign (MAHA). These results utilize optical spectra obtained with the Long Slit Spectrograph on the Wyoming Infrared 2.3m Telescope between 2017 November---2023 May. MAHA combines long-duration monitoring with high cadence. We report results from multiple observing seasons for 9 of the 14 objects. These results include H$\beta$ time lags, supermassive black hole masses, and velocity-resolved time lags. The velocity-resolved lags allow us to investigate the kinematics of the broad-line region. 
\end{abstract}


\section{Introduction} \label{sec:intro} 
Reverberation mapping (RM) is a spectroscopic monitoring technique that measures the masses of the supermassive black holes (SMBHs) fueling active galactic nuclei (AGNs) \citep[e.g.,][]{Blandford1982,Peterson1993,Cackett2021}. RM takes advantage of two defining observational signatures of AGNs: 1) variability on short time scales (days to months) and 2) broad emission lines in their UV and optical spectra. These broad emission lines (BELs) arise from gas in the broad-line region (BLR), photoionized by the ionizing continuum. The BLR flux echoes the ionizing continuum flux density variations some time later due to the finite speed of light. RM uses spectroscopic monitoring to measure the delayed response of the BLR, $\tau_{\rm BLR}$, which is converted to an average radius of the BLR, R$_{\rm BLR}$, by multiplying by the speed of light. Assuming virialized motion of the BLR gas, we can use $\tau_{\rm BLR}$ along with the BLR velocity, measured from the width of a BEL, to estimate the SMBH mass. 

Through great observational effort over the past three decades, more than 100 AGNs' SMBH masses have been measured with RM \citep[e.g.,][]{Barth2015,Bentz2009,Chen2023,Denney2010,Du2018,Edelson2015,Fausnaugh2016,Grier2012,Grier2017,Kaspi2000,Kaspi2007,Lu2022,Malik2023,Pandey2022,Peterson2004,Shen2015,Xiao2011}. RM anchors the empirical relationship between the BLR radius and AGN luminosity \citep{Kaspi2000,Bentz2013,Du2019} that allows for single-epoch mass estimation.

In addition to determining SMBH masses, RM can also reveal information about the BLR geometry and kinematics. With sufficiently high cadence and signal-to-noise (S/N) data, we can measure $\tau_{\rm BLR}$ in velocity bins across an emission line. This allows for broad conclusions as to the  dominant motion of the BLR gas (e.g., disk-like, outflow, or inflow) \citep{Horne2004}. There are $\sim$45 objects with velocity-resolved time lags \citep{MAHA3,Bentz2008,Bentz2009,Bentz2021a,Bentz2022,Bentz2023,MAHA2,Denney2009c,Denney2010,DeRosa2018,Du2016b,MAHA1,Hu2020a,Li2022,Lu2016,Valenti2015,U2022}. Advanced techniques such as two-dimensional velocity delay maps (VDMs) \citep{Horne2004} and dynamical modeling of the BLR \citep{Pancoast2014,Li2013} can reveal even more information about the BLR and provide a check on the fidelity of RM masses. 

Monitoring AGNs with H$\beta$ Asymmetry (MAHA) is an ongoing long-term RM campaign of objects with asymmetric H$\beta$ emission lines. MAHA began in 2016 and has observed $\sim$100 objects for one or more seasons at different priority levels. Our primary targets have high-cadence, long-duration light curves that allow for not only SMBH mass estimations, but also advanced data analysis techniques such as measuring the line response in multiple velocity bins or dynamical modeling of the BLR gas. For more information on the MAHA campaign, see the first three MAHA papers, hereafter referred to as MAHA I \citep{MAHA1}, MAHA II \citep{MAHA2}, and MAHA III \citep{MAHA3} as well as the following papers utilizing MAHA data: \citet{Kara2021,Oknyansky2021,Chen2023,Oknyansky2023}. We provide here a short summary of published MAHA results: MAHA I presented time lags, corresponding black hole masses, and velocity-resolved time lags for 3C 120, Ark 120, Mrk 6, and SBS 1518+ 593; MAHA II presented time lags, corresponding black hole masses, and velocity-resolved time lags for Mrk 79, NGC 3227, and Mrk 841. MAHA II also presented velocity-delay maps for Mrk 79 and NGC 3227; MAHA III presented time lags, corresponding black hole masses, and velocity-resolved time lags for 15 Palomar-Green quasars.
In this paper, we present 14 MAHA objects without prior RM results. There are good results for multiple seasons for 9 of the objects, 
and we show all the seasons here. We present velocity-resolved lags for all objects and discuss implications for the kinematics of the individual BLRs. 

Section~\ref{sec:obs} describes our observations and data reduction as well as the construction of our light curves. Section~\ref{sec:analysis} describes our analysis including time-lag measurements, line-width measurements, black hole mass calculations, and velocity-resolved time lag measurements. Section~\ref{sec:dis} contains the discussion of the results for each object. Finally, Section~\ref{sec:con} summarizes our results.

\section{Observations and Data Reduction} \label{sec:obs}
In this paper, we present results from reverberation mapping of 14 MAHA AGNs that have not previously had successful time lags published. For more information about the MAHA target list as a whole, including sample selection criterion, see MAHA I, although we note that our targets have evolved over time. In short, we prefer bright, northern objects with asymmetric H$\beta$ profiles as measured by the dimensionless asymmetry parameter

    \begin{equation}
        A = \frac{\lambda_c(3/4) - \lambda_c(1/4)}{\Delta\lambda(1/2)}
    \end{equation}

where $\lambda_c$(3/4) and $\lambda_c$(1/4) are the central wavelengths at 3/4 and 1/4 of the line peak and $\Delta\lambda$(1/2) is the FWHM of the line \citep{DeRobertis1985,BorosonGreen1992,Brotherton1996}. Blue asymmetries are positive and red are negative. 
Usually targets have been selected based on published or publicly available spectra showing asymmetric H$\beta$, but the profile may have varied by the time we began observing it.
Target information, including coordinates and redshifts from NED,\footnote{The NASA/IPAC Extragalactic Database (NED) is funded by the National Aeronautics and Space Administration and operated by the California Institute of Technology.} are listed in  Table~\ref{tab:targets}. We next describe our observations and data reduction.  

\begin{deluxetable*}{lccc}
\tablecaption{Featured Targets\label{tab:targets}}
\tablewidth{0pt}
\tablehead{
\colhead{Object} & \colhead{$\alpha 2000$} & \colhead{$\delta 2000$} & \colhead{Redshift}\\
\colhead{} & \colhead{(hr min sec)} & \colhead{($^{\circ}$ \textquotesingle \hspace{1mm}\textquotesingle\textquotesingle)} & \colhead{} \\
\colhead{(1)} & \colhead{(2)} & \colhead{(3)} & \colhead{(4)}
}

\startdata
1ES 0206+522 & 02 09 37.4 & +52 26 40 & 0.0492 \\
Mrk 1040 & 02 28 14.5 & +31 18 42 & 0.0166 \\
Mrk 618 & 04 36 22.2 & -10 22 34 & 0.0356 \\
MCG -02-14-009 & 05 16 21.2 & --10 33 41 & 0.0285 \\
IRAS 05589+2828 & 06 02 10.7 & +28 28 22 & 0.0294 \\
Mrk 715 & 10 04 47.6 & +14 46 46 & 0.0845 \\
SBS 1136+594 & 11 39 09.0 & +59 11 55 & 0.0612 \\
VIII Zw 233 & 13 05 34.5 & +18 19 33 & 0.1182 \\
Mrk 813 & 14 27 25.1 & +19 49 52 & 0.1099 \\
SDSS J145307.92+255433.0 & 14 53 07.9 & +25 54 33 & 0.0485 \\
SDSS J152139.66+033729.2 & 15 21 39.7 & +03 37 29 & 0.1260 \\
2MASX J21090996-0940147 & 21 09 10.0 & -09 40 15 & 0.0270 \\
PG 2304+042 & 23 07 03.0 & +04 32 57 & 0.0466 \\
NGC 7603 & 23 18 56.6 & +00 14 38 & 0.0288 \\ 
\enddata
\tablecomments{Object information. All coordinates and redshifts from NED except for 1ES 0206+522 for which we used coordinates from SIMBAD and the redshift from NED.}
\end{deluxetable*}

\subsection{Spectrophotometry} \label{subsec:spec}
We conducted our observations using the Wyoming Infrared Observatory's (WIRO) 2.3m telescope with the Long Slit Spectrograph primarily in remote operation mode \citep{Findlay2016}. We used a 900 line mm$^{-1}$ grating which provides a wavelength range of 4000---7000 \r{A} and a dispersion of 1.49 \r{A} pixel$^{-1}$. We used a 5$''$-wide slit oriented north-south. We observed at least one spectrophotometric standard star each night for flux calibration (either BD+28d4211, Feige 34, G191B2B, or HZ 44). We observed CuAr lamps for wavelength calibration and a uniformly lit white square for flat fields. 

We reduced the data with IRAF v2.16
using standard techniques and a uniform aperture of $\pm$ 6$\farcs$8 and background windows of 7$\farcs$6---15$''$.1 on both sides of the objects' nuclear emission. 

There are sometimes flux density variations between spectra due to clouds and slit losses. In order to obtain accurate photometry, we performed additional flux calibrations using the [O III] $\lambda$5007 line. This [O III] line arises from the narrow line region (NLR) of the AGN which varies slowly, on the order of tens of years or longer, making them appropriate for use in an observation campaign of months \citep{Peterson2013}. We used the calibration technique of \citet{vanGronigen1992} with minor modifications. Briefly, our wide slit (5$''$) is larger than the average FWHM of the seeing of our observations (2$''$--3$''$), and variable seeing between exposures leads to changes in the line point spread function (PSF). In addition, telescope tracking errors can lead the line PSF to deviate from a Gaussian. To account for these effects we convolved the [O III] profiles with a double Gaussian that fits the broadest [O III] profile observed for each object. This smooths the spectra and minimizes the effect of variable seeing and tracking inaccuracies. We then scaled each exposure of an object to match its standard O [III] flux. Standard O [III] fluxes are obtained from spectra taken in photometric conditions. For more details, see discussion in MAHA I. Our standard [O III] fluxes are listed in Table~\ref{tab:OIII}. 

\begin{figure*}[t]
        \includegraphics[width=1\textwidth]{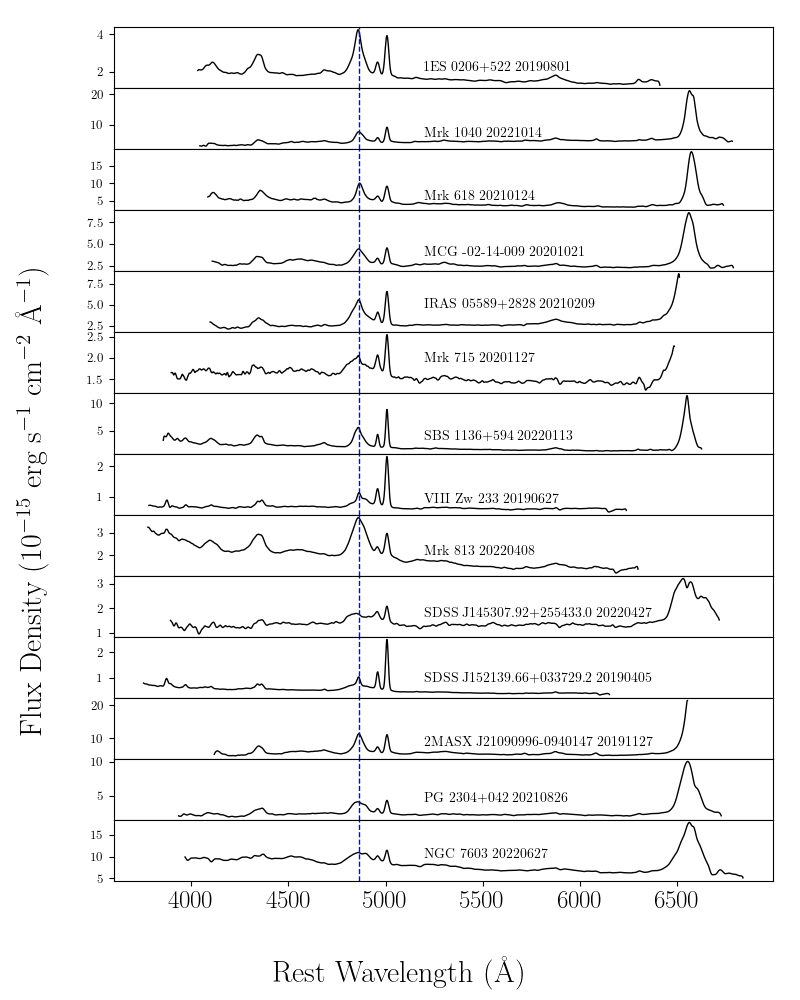}
        \caption{Randomly selected typical WIRO spectra for each object analyzed in this paper. The spectra are reduced and flux calibrated as discussed in Section~\ref{subsec:lightcurves}. Flux density units are 10$^{-15}$ erg s$^{-1}$ cm$^{-2}$ \r{A}$^{-1}$. Wavelengths are in the rest frame. The vertical dashed blue line indicates the center of the H$\beta$ line. Each spectrum is labeled with the name of the object and civil date observed.}
        \label{figure:random_spec}
        \end{figure*}

We produced the final flux-calibrated spectra by averaging the exposures (appropriately noise weighted) from a single night. See Figure~\ref{figure:random_spec} for examples of typical MAHA spectra. 

In order to investigate the general H$\beta$ profile and measure line asymmetries and widths, we calculated mean and rms spectra using
    \begin{equation}
        \bar{F}_{\lambda} = \frac{1}{N}\sum_{i=1}^{N}F_{\lambda}^i
    \end{equation}
and
    \begin{equation}
        S_{\lambda} = \left[ \frac{1}{N}\sum_{i=1}^{N}(F_{\lambda}^i - \bar{F}_{\lambda})^2 \right]^{1/2}
    \end{equation}
respectively. 

We present mean and rms spectra for each object in Figure~\ref{figure:1es_lc}.
Note that the [O III] lines have essentially disappeared in the rms spectra, indicating data uniformity and consistency in our flux calibration.
 
\begin{deluxetable*}{lcccc}
\tablecaption{Measurement Windows and [O III] Standard Fluxes\label{tab:OIII}}
\tablewidth{0pt}
\tablehead{
\colhead{Object} & \colhead{F$_{[\textrm{O III}]}$} & \colhead{Continuum (blue)} & \colhead{Continuum (red)} & \colhead{H$\beta$} \\
\colhead{} & \colhead{(10$^{-13}$ erg s$^{-1}$ cm$^{-2}$)} & \colhead{(\r{A})} & \colhead{(\r{A})} & \colhead{(\r{A})} \\
\colhead{(1)} & \colhead{(2)} & \colhead{(3)} & \colhead{(4)} & \colhead{(5)}
}

\startdata
1ES 0206+522 & 0.68 & 4790-4800 & 5037-5057 & 4800-4936 \\
Mrk 1040 & 0.96 & 4786-4796 & 4975-4985 & 4796-4920 \\
Mrk 618 & 1.1 & 4782-4802 & 4974-4984 & 4802-4935 \\
MCG -02-14-009 & 0.43 & 4751-4771 & 4974-4984 & 4771-4930 \\
IRAS 05589+2828 & 0.86 & 4735-4755 & 4985-4995 & 4755-4975 \\
Mrk 715 & 0.19 & 4720-4740 & 5031-5051 & 4740-5031\\
SBS 1136+594 & 1.0 & 4782-4800 & 4973-4987 & 4800-4935 \\
VIII Zw 233 & 0.36 & 4773-4789 & 4930-4949 & 4789-4930 \\
Mrk 813 & 0.25 & 4745-4765 & 4974-4982 & 4795-4945 \\
SDSS J145307.92+255433.0.0 & 0.12 & 4750-4770 & 5030-5040 & 4770-5030 \\
SDSS J152139.66+033729.2 & 0.37 & 4712-4732 & 5075-5125 & 4732-5040 \\
2MASX J21090996-0940147 & 1.2 & 4780-4800 & 4922-4942 & 4800-4922\\
PG 2304+042 & 0.41 & 4777-4797 & 4975-4985 & 4797-4933 \\
NGC 7603 & 0.60 & 4734-4754 & 5034-5044 & 4754-5034 
\enddata
\tablecomments{The second column contains the O [III] $\lambda$5007 standard fluxes used for flux calibration. The third and fourth columns contain the windows on the blue and red side of H$\beta$ used for fitting a linear continuum underneath the line. The fifth column contains the window used for integrating across the H$\beta$ profile. All wavelengths are in the rest frame.}
\end{deluxetable*}

\subsection{WIRO Light Curves} \label{subsec:lightcurves}
Before constructing our light curves, we corrected each epoch of our calibrated spectra for Galactic reddening. We used the \citet{CCM1989} extinction law implemented by the Python package dust\_extinction and the \citet{SchlaflyFinkbeiner2011} R$_{\rm{V}}$=3.1 and A$_{\rm{V}}$ values for each object taken from NED.

We measured the 5100 \r{A} continuum flux densities from our calibrated spectra by averaging the flux density spanning the wavelength range of 5075 to 5125 \r{A} in the rest frame. We note that we do not remove host-galaxy starlight from our continuum flux densities.

There are two commonly used ways to measure emission line fluxes: direct integration \citep[e.g.,][]{Peterson1998,Kaspi2000,Bentz2009,Grier2012,Du2014} or spectral fitting \citep[e.g.,][]{Barth2013,Hu2015}. Direct integration involves fitting a linear continuum underneath the H$\beta$ profile, subtracting that continuum from the line profile, then integrating across the line profile. Spectral fitting involves detailed fitting of the various components of the spectrum to isolate the broad emission line. We followed MAHA I in using direct integration. To fit a linear continuum underneath the H$\beta$ profile, we linearly interpolated across two emission line-free windows on either side of the line profile. We used the rms spectrum to select the emission-line-free windows, choosing minimally varying windows to the left and right of the H$\beta$ emission line. We also used the rms spectrum to select a H$\beta$ integration window that contains the line as shown in the rms spectrum. Table~\ref{tab:OIII} lists H$\beta$ integration windows for each object.

The broad H$\beta$ emission contains flux contamination from the narrow H$\beta$ component and, in some objects, the [O III] $\lambda\lambda$4959 5007 doublet. We isolated the broad component of the H$\beta$ emission line by subtracting any narrow-line component from our measured H$\beta$ fluxes. We began by fitting the [O III] doublet and narrow H$\beta$ in each individual spectrum (for the fitting details see Section~\ref{subsec:line_width}). We then integrated the fits to obtain the narrow-line fluxes. Finally we subtracted from each epoch the narrow-line fluxes included in each object's H$\beta$ profile as determined from their rms profiles.

In addition to narrow-line contamination, the broad H$\beta$ emission can also contain flux contamination from Fe II emission. Only two of our 14 objects show moderate optical Fe II emission: Mrk 813 and NGC 7603. Fe II varies on similar time scales as H$\beta$ \citep{Hu2015}, consequently it should not strongly affect the H$\beta$ lag measurement. Fe II emission is also relatively weak in the H$\beta$ region, though it is stronger in the region of the O [III] doublet, making Fe II contamination a concern only when the broad H$\beta$ emission is blended with O [III]. Mrk 813 has H$\beta$ emission that is not blended with O [III] (see Figure~\ref{figure:1es_lc}) and NGC 7603 has H$\beta$ emission that is blended with O [III] (see Figure~\ref{figure:1es_lc}). We tried changing the H$\beta$ integration limits for NGC 7603 in order to remove the O [III], and as a result the Fe II, emission, from the H$\beta$ integration range. This results in a lag measurement that is consistent with the lag measured using the full H$\beta$ integration range, to within the uncertainties (see Table~\ref{tab:OIII} for the H$\beta$ integration range). This indicates that in NGC 7603 the Fe II emission that is blended with the broad H$\beta$ emission does not significantly affect the time lag. We therefore elected to keep our full H$\beta$ integration range for NGC 7603 and ignore the potential Fe II contamination.

The flux density and flux uncertainties in the light curve contain two components: 1) Poisson noise, and 2) random uncertainty from tracking error and variable observing conditions in different exposures of an object during the same night. We estimated the random uncertainty using the scatter of the flux density between consecutive exposures on the same night between $\sim$4100---5700 \r{A}. We added these two sources of uncertainty in quadrature. We emphasize here that our error budget does not include the systematic uncertainty from our narrow-line model assumptions used to subtract the narrow-line contribution from H$\beta$. We also do not include uncertainty contributed by the narrow-line subtraction itself. Therefore our total H$\beta$ flux uncertainties are likely underestimated. We plotted the total uncertainty as error bars in our light curves.

\subsection{Supplemental Continuum Light Curves from Zwicky Transient Facility Photometry} \label{subsec:ZTF}
We followed MAHA I in supplementing our continuum light curves with survey photometry when available and of sufficiently small uncertainties. Contributions from optical emission lines (other than H$\alpha$) contaminating broadband filters is generally weak, making broadband photometry usually acceptable for continuum light curves. The additional data points provide improved temporal coverage and cadence. 

The Zwicky Transient Facility is an all-northern-sky survey using the Palomar 48-inch Schmidt Telescope at Palomar Observatory. The survey goes to median depths of g $\sim$ 20.8 and r $\sim$ 20.6 mag. The survey began in March of 2018 and builds upon the predecessor survey, the Palomar Transient Factory \citep{Masci2019,ztf_doi}. When available, we used ZTF g-band photometry to supplement our 5100 \r{A} light curves. We chose g-band over r-band photometry as g-band contains (rest frame) 5100 \r{A} for objects with z $\leq$ 0.11, making it the appropriate choice for our sample \citep{Bellm2019}. We utilized the ztfquery package \citep{Rigault2018} in fetching the ZTF light curves. 

It is necessary to scale the ZTF photometry to account for differing host galaxy flux density contributions due to aperture size differences between ZTF and WIRO. The scaling is a simple linear shift following

    \begin{equation}
        F_{5100} = a + bF_{ZTF}.
    \end{equation}

To find the optimal $a$ and $b$ parameters we used the Bayesian-statistics-based package PyCali;\footnote{PyCALI is available at: \href{https://github.com/LiyrAstroph/PyCALI}{https://github.com/LiyrAstroph/PyCALI}.} PyCali assumes the light curve can be described by a damped random walk model \citep{Li2014}. We then used the best-fit parameters to scale the ZTF flux densities. We note that PyCali takes into account the uncertainty on $a$ and $b$ in the final uncertainty of the scaled ZTF flux densities. Finally, we combined data points from the same night using a weighted average. Visual inspection of the light curves show that the inter-calibrated continuum data agree well.

\begin{figure*}[t]
        \includegraphics[width=1\textwidth]{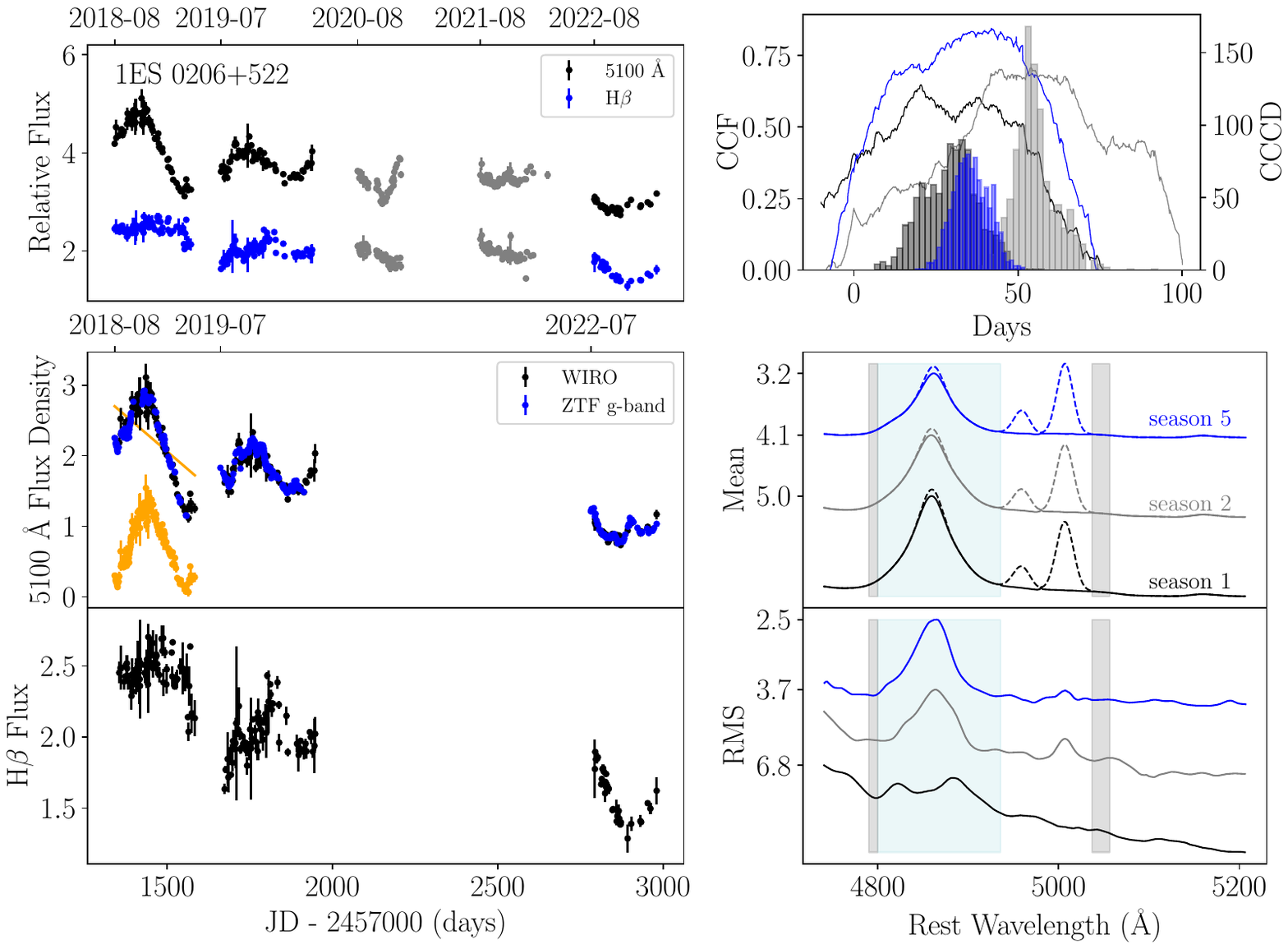}
        \caption{Time-series analysis of 1ES 0206+522. The top left panel shows the WIRO continuum and H$\beta$ light curves for each season in arbitrary flux density or flux units respectively, shifted relative to each other so as to both display on the plot. The grayed out seasons do not give a reliable result. The middle left panel shows combined WIRO and ZTF continuum light curves for each analyzed season in units of 10$^{-15}$ erg s$^{-1}$ cm$^{-2}$ \r{A}$^{-1}$. The orange line in season 1 shows the detrending line fit to the continuum. The orange season 1 light curve shows the detrended light curve. The bottom left panel are H$\beta$ light curves for each analyzed season in units of 10$^{-13}$ erg s$^{-1}$ cm$^{-2}$. The top right panel are CCFs and CCCDs in the rest frame for each analyzed season (the season 1 CCF/CCCD is measured using the detrended continuum light curve). The middle right panel shows the AGN power continuum-subtracted mean spectrum for each analyzed season in units of 10$^{-16}$ erg s$^{-1}$ cm$^{-2}$ \r{A}$^{-1}$. The dashed line shows the [O III] doublet and H$\beta$ narrow-line components and the solid line shows the H$\beta$ broad component. The bottom right panel are the rms spectra for each analyzed season in units of 10$^{-16}$ erg s$^{-1}$ cm$^{-2}$ \r{A}$^{-1}$. Both the mean and rms spectra are shifted relative to each other in order to display all seasons on one plot. The y-axis ticks and tick labels on the mean and rms spectra plots correspond to the maximum flux density value of each spectrum shown. The blue region is the H$\beta$ integration range and the grey regions are the line-free continuum regions used to fit the linear continuum under H$\beta$. All wavelengths are in the rest frame. The colors in the right panels correspond to the different seasons as defined in the middle right panel.}
        \label{figure:1es_lc}
        \end{figure*}

\begin{figure*}[t]
        \includegraphics[width=1\textwidth]{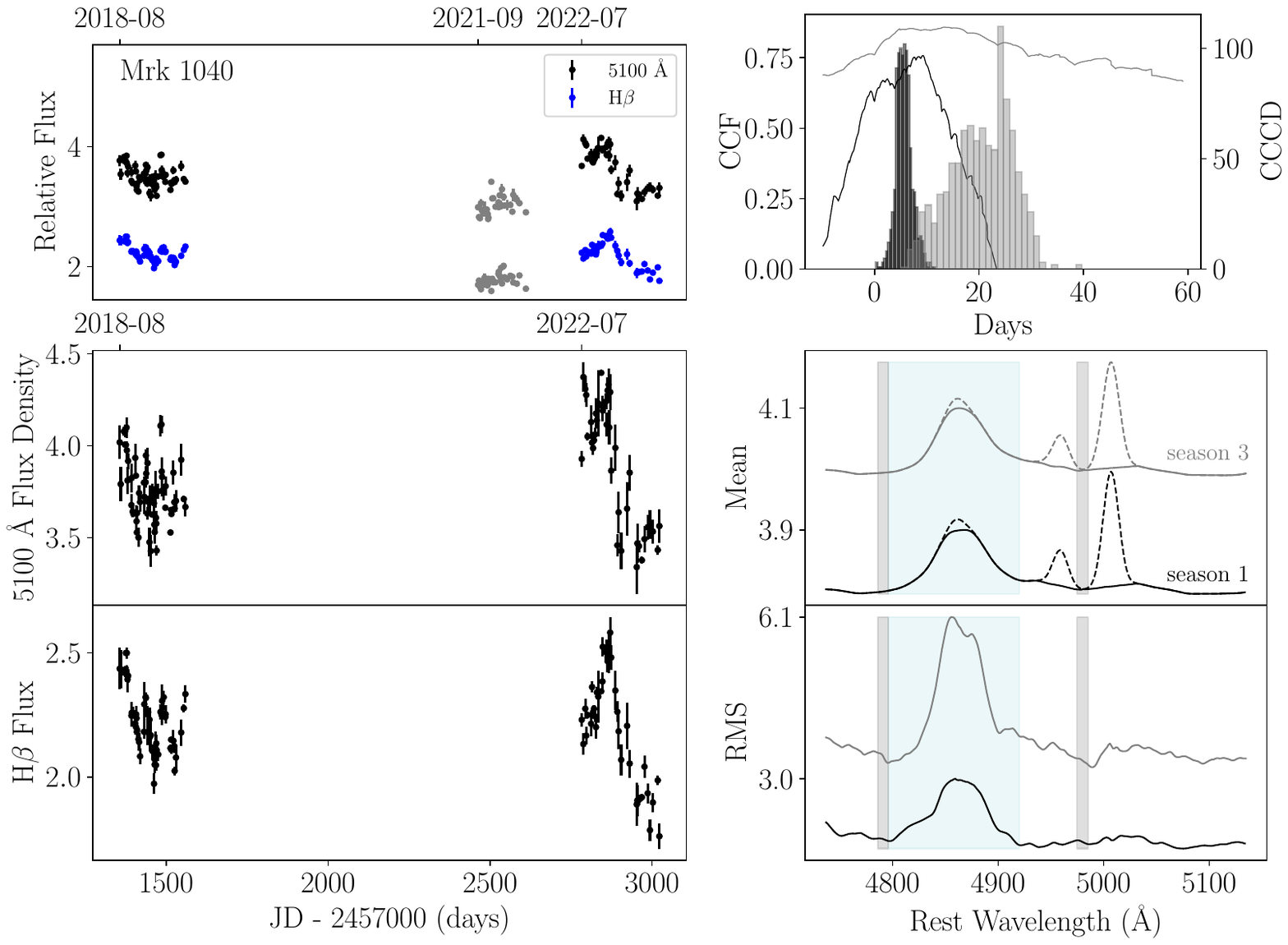}
        \caption{Time series analysis of Mrk 1040. The meanings of the panels are the same as Figure~\ref{figure:1es_lc}.}
        \label{figure:mrk1040_lc}
        \end{figure*}

\begin{figure*}[t]
        \includegraphics[width=1\textwidth]{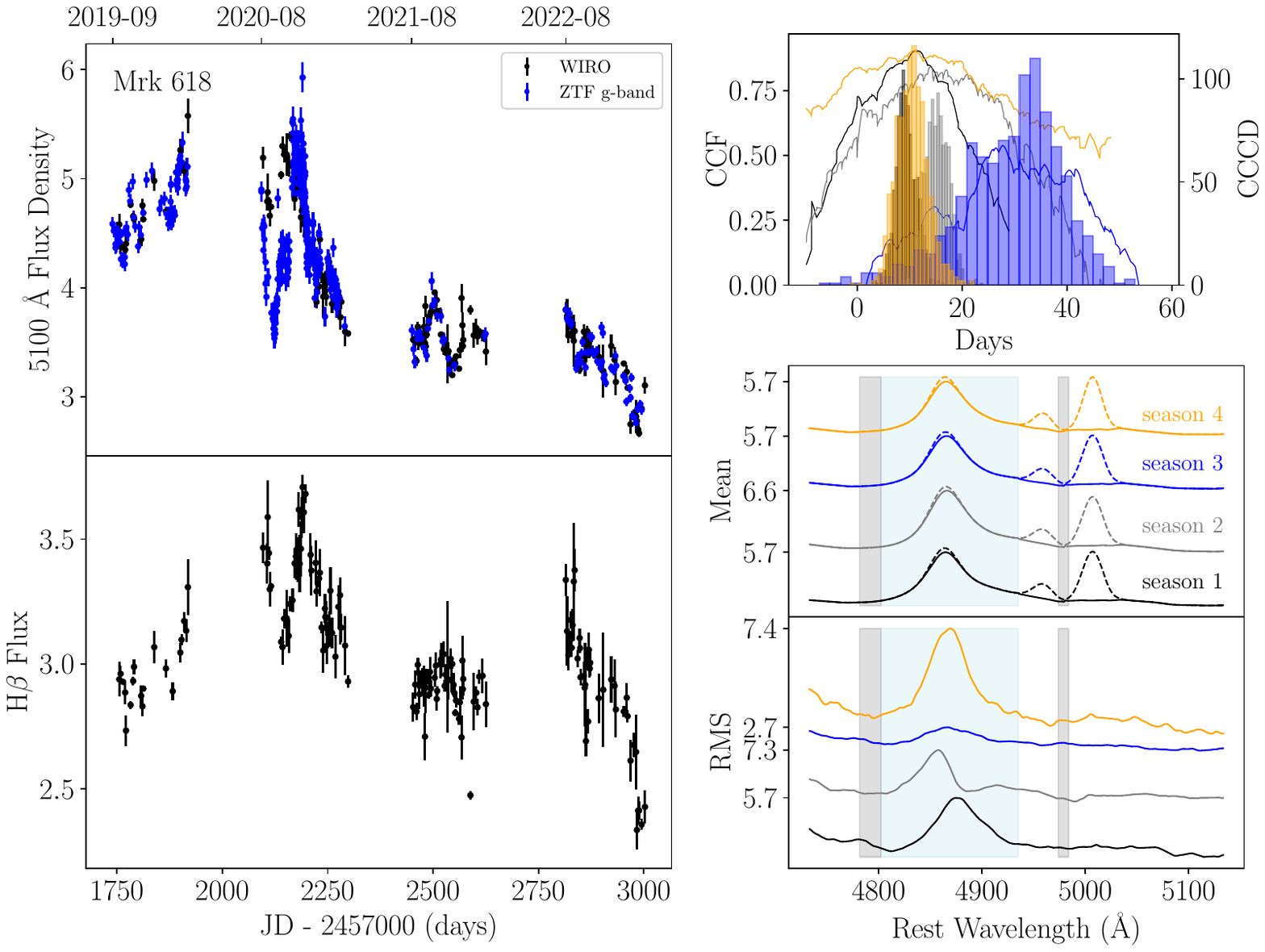}
        \caption{Time series analysis of Mrk 618. There are only two left panels as we analyze all WIRO seasons for this object. The meanings of the panels are the same as Figure~\ref{figure:1es_lc}.}
        \label{figure:mrk618_lc}
        \end{figure*}

\begin{figure*}[t]
        \includegraphics[width=1\textwidth]{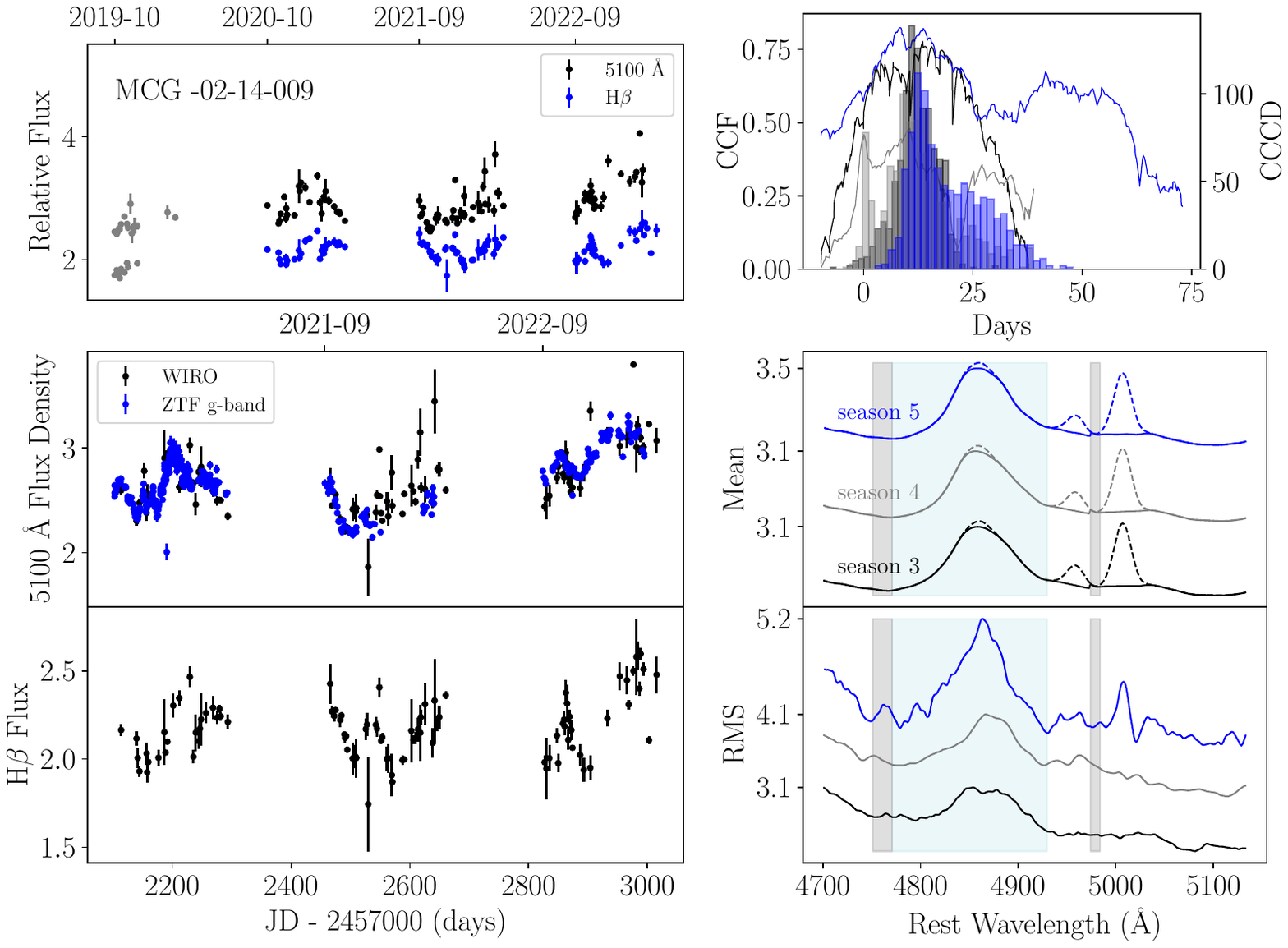}
        \caption{Time series analysis of MCG -02-14-009. The meanings of the panels are the same as Figure~\ref{figure:1es_lc}.}
        \label{figure:mcg_lc}
        \end{figure*}

\begin{figure*}[t]
        \includegraphics[width=1\textwidth]{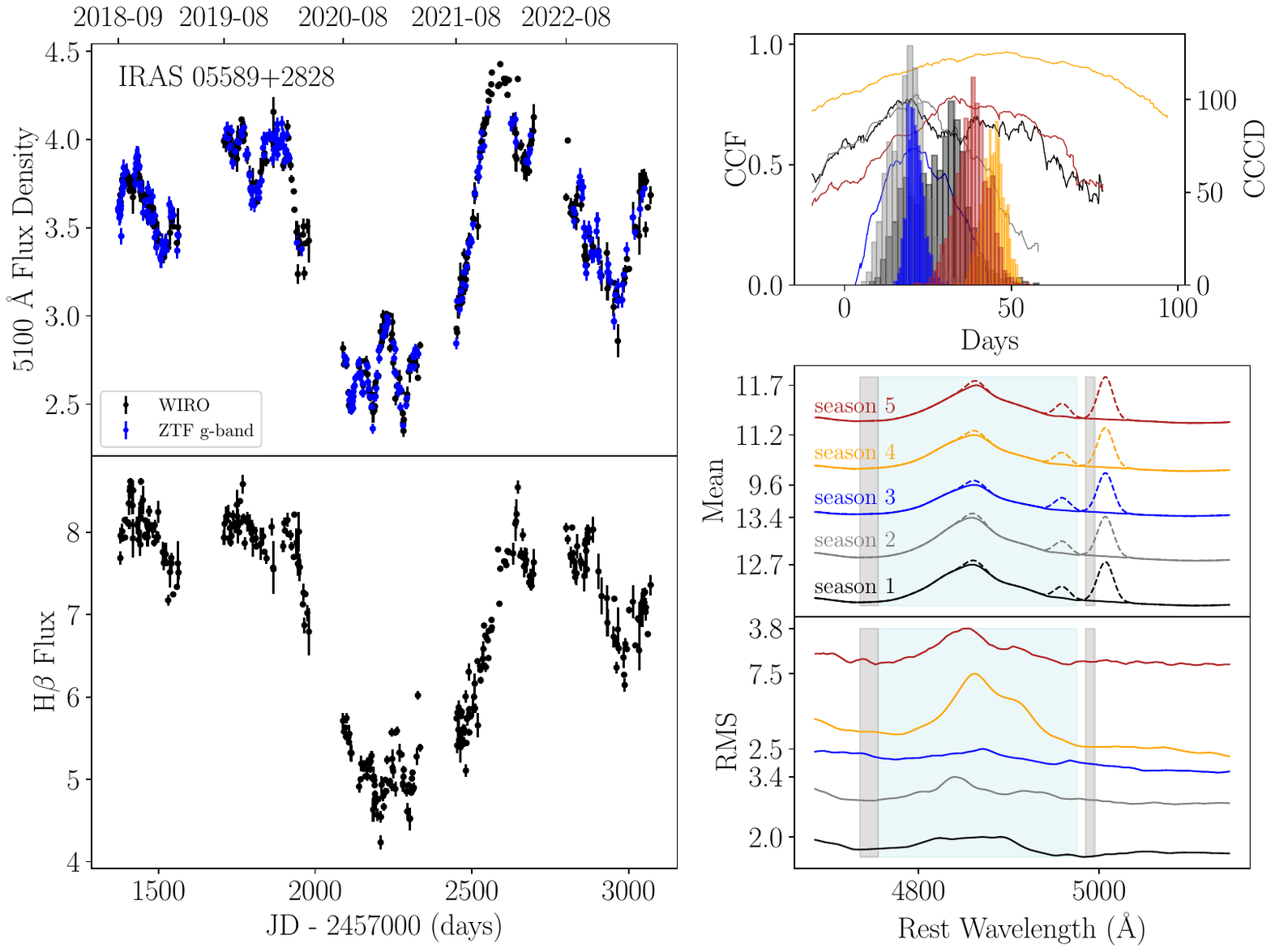}
        \caption{Time series analysis of IRAS 05589+2828. There are only two left panels as we analyze all WIRO seasons for this object. The meanings of the panels are the same as Figure~\ref{figure:1es_lc}.}
        \label{figure:iras_lc}
        \end{figure*}

\begin{figure*}[t]
        \includegraphics[width=1\textwidth]{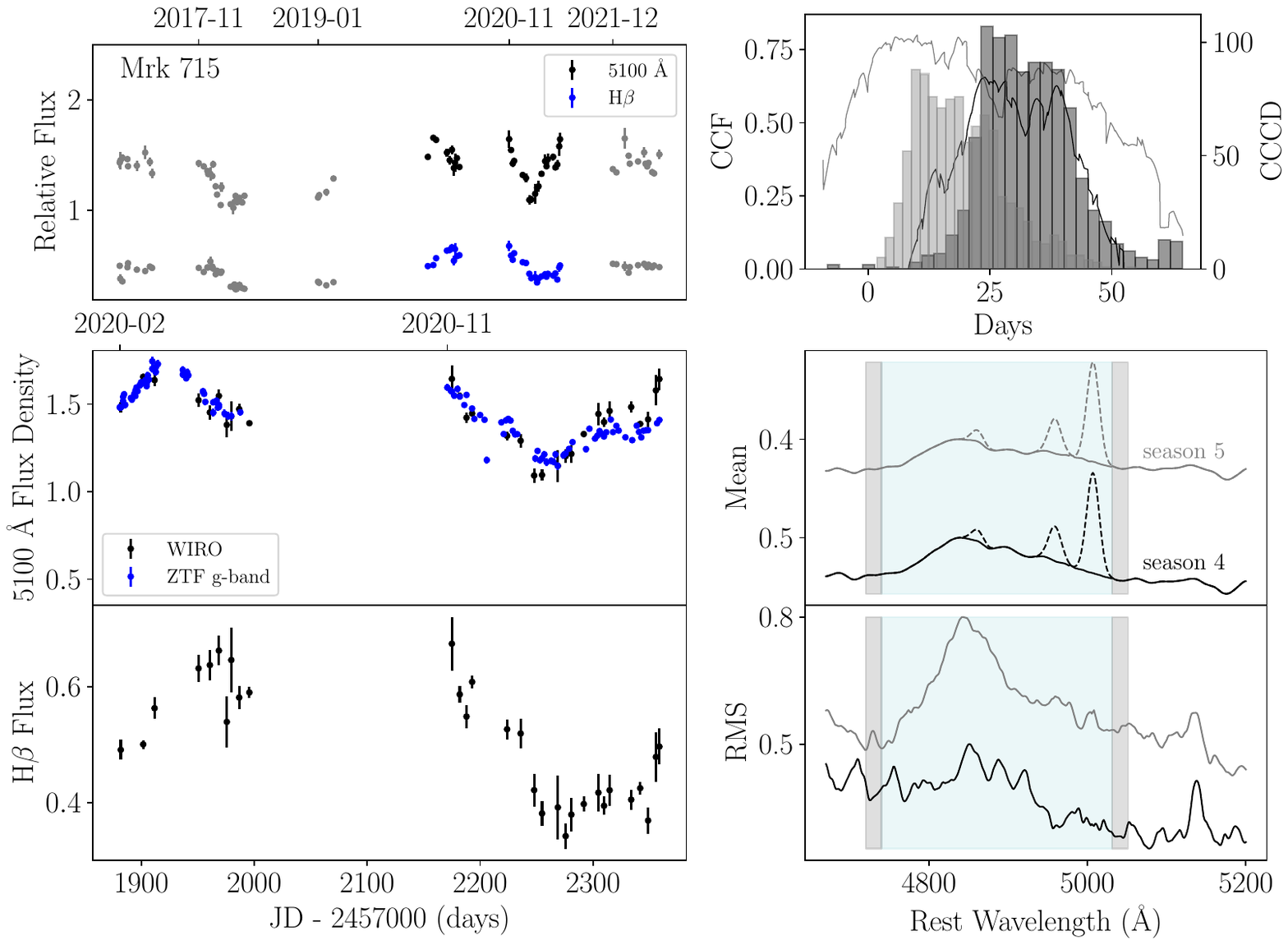}
        \caption{Time-series analysis of Mrk 715. The meanings of the panels are the same as Figure~\ref{figure:1es_lc}.}
        \label{figure:sdss1004_lc}
        \end{figure*}

\begin{figure*}[t]
        \includegraphics[width=1\textwidth]{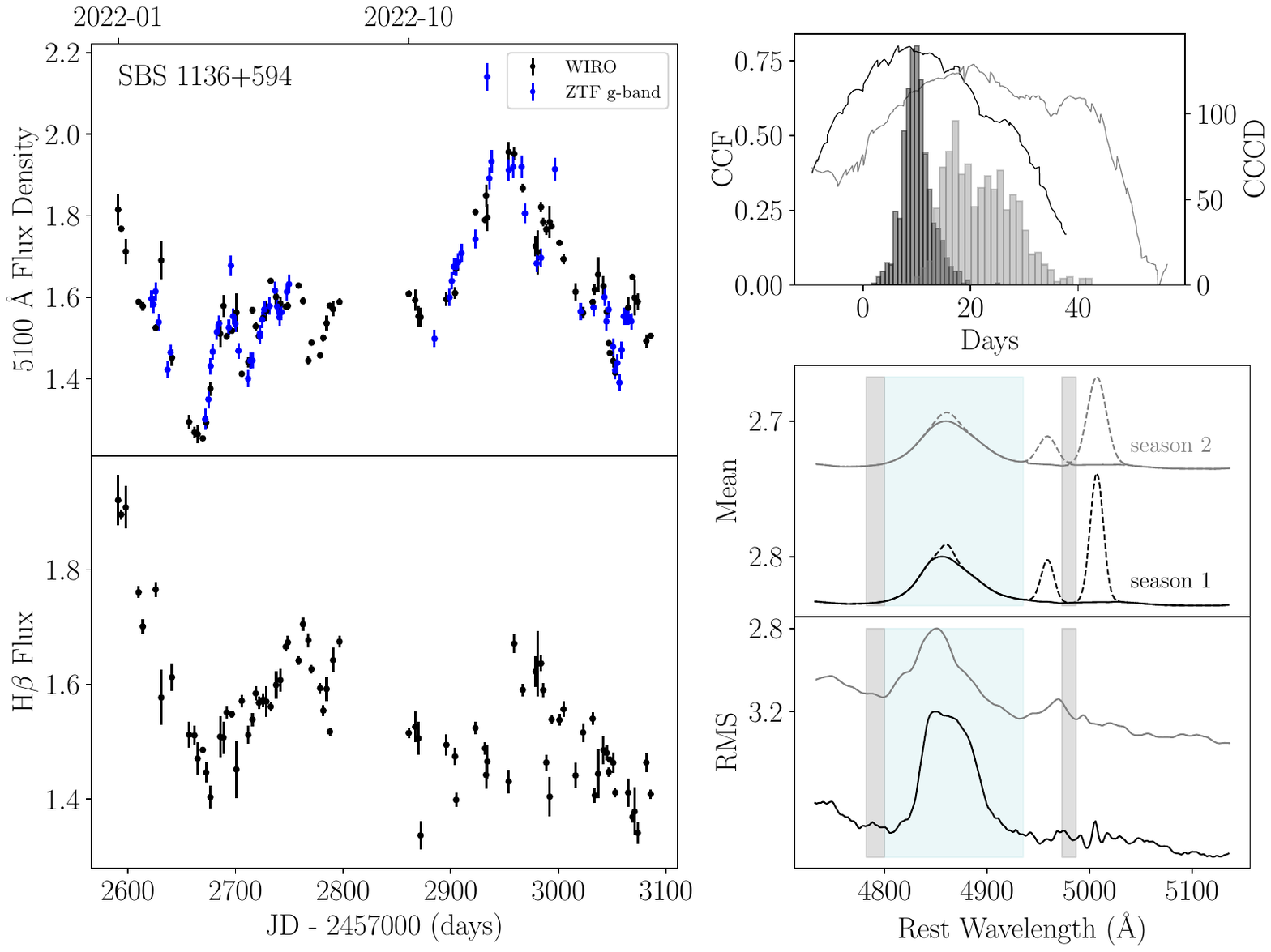}
        \caption{Time-series analysis of SBS 1136+594. There are only two left panels as we analyze all WIRO seasons for this object. The meanings of the panels are the same as Figure~\ref{figure:1es_lc}.}
        \label{figure:sbs_lc}
        \end{figure*}

\begin{figure*}[t]
        \includegraphics[width=1\textwidth]{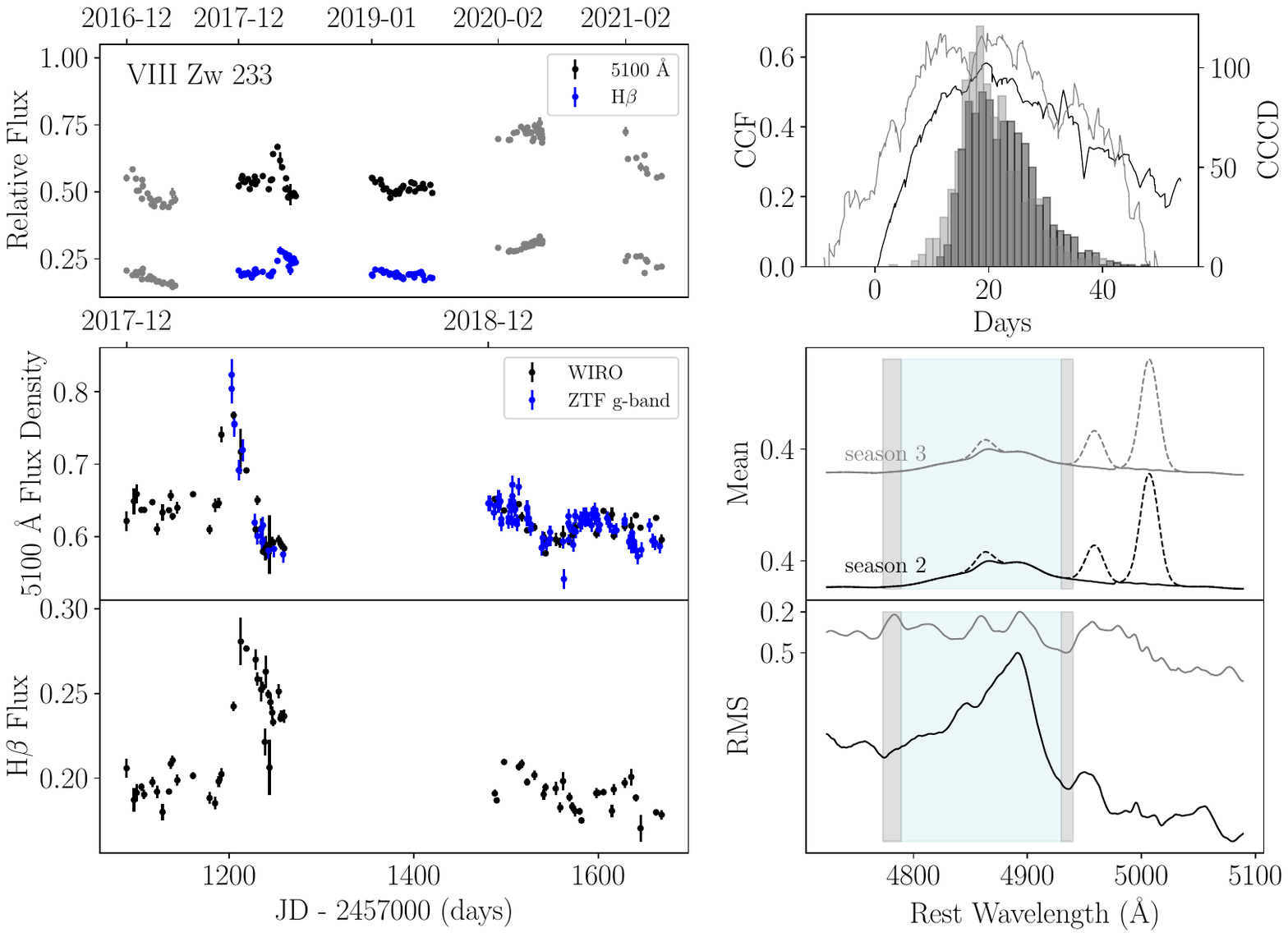}
       \caption{Time-series analysis of VIII Zw 233. The meanings of the panels are the same as Figure~\ref{figure:1es_lc}.}
        \label{figure:viiizw233_lc}
        \end{figure*}

\begin{figure*}[t]
        \includegraphics[width=1\textwidth]{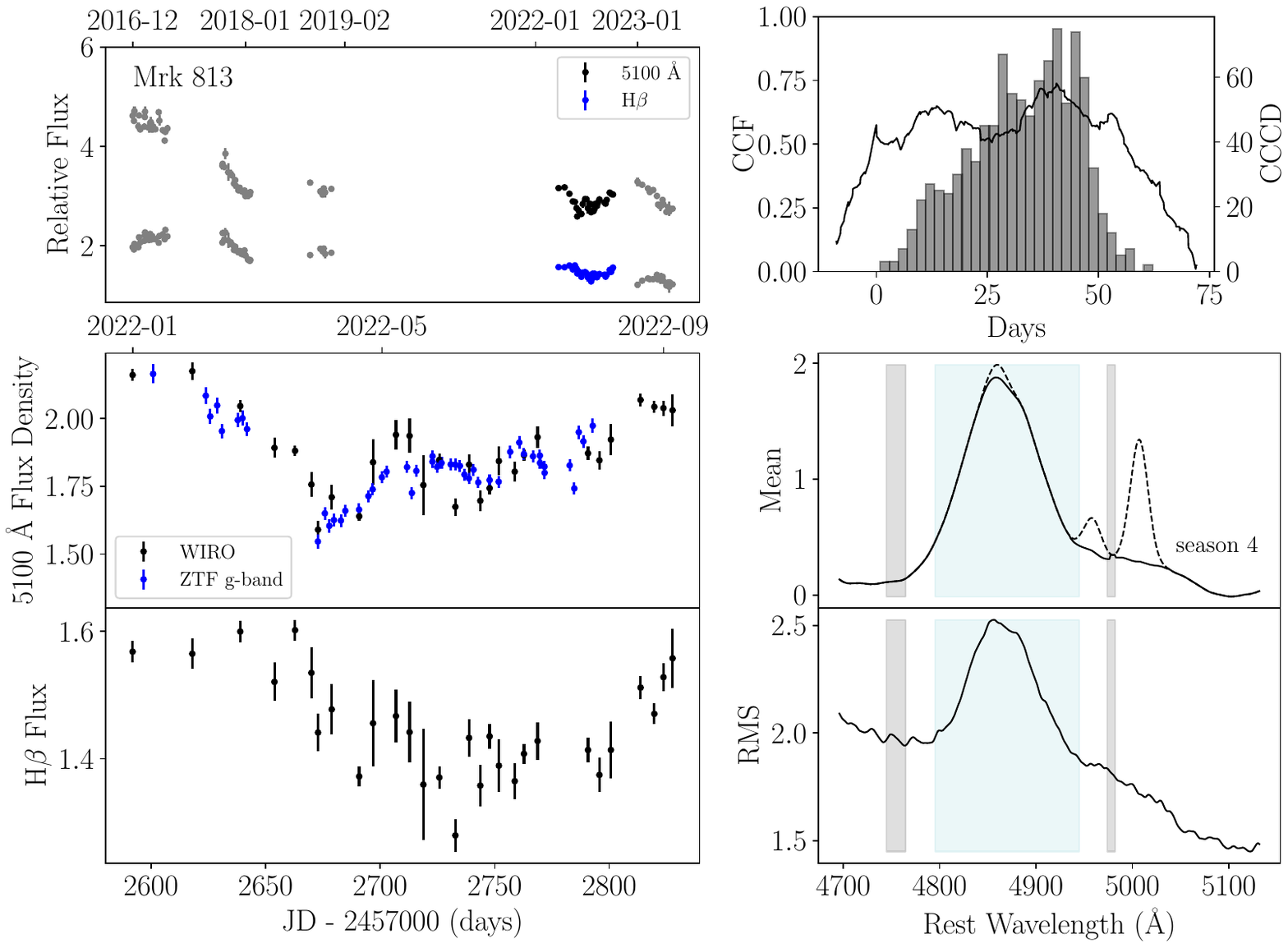}
        \caption{Time-series analysis of Mrk 813. The meanings of the panels are the same as Figure~\ref{figure:1es_lc}.}
        \label{figure:mrk813_lc}
        \end{figure*}

\begin{figure*}[t]
        \includegraphics[width=1\textwidth]{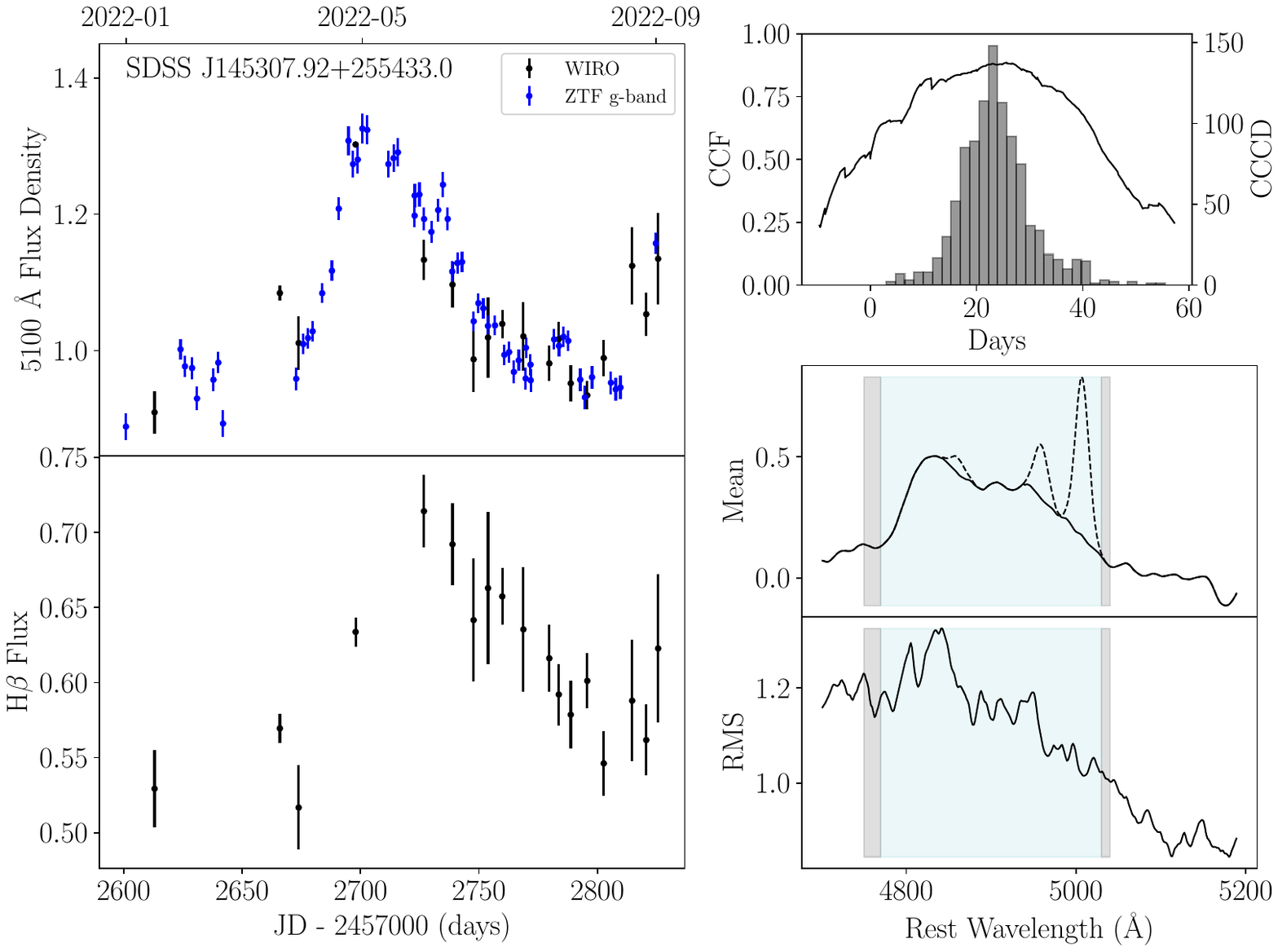}
        \caption{Time-series analysis of SDSS J145307.92+255433.0. There are only two left panels as we analyze all WIRO seasons for this object. The meanings of the panels are the same as Figure~\ref{figure:1es_lc}.}
        \label{figure:sdssj1453_lc}
        \end{figure*}

\begin{figure*}[t]
        \includegraphics[width=1\textwidth]{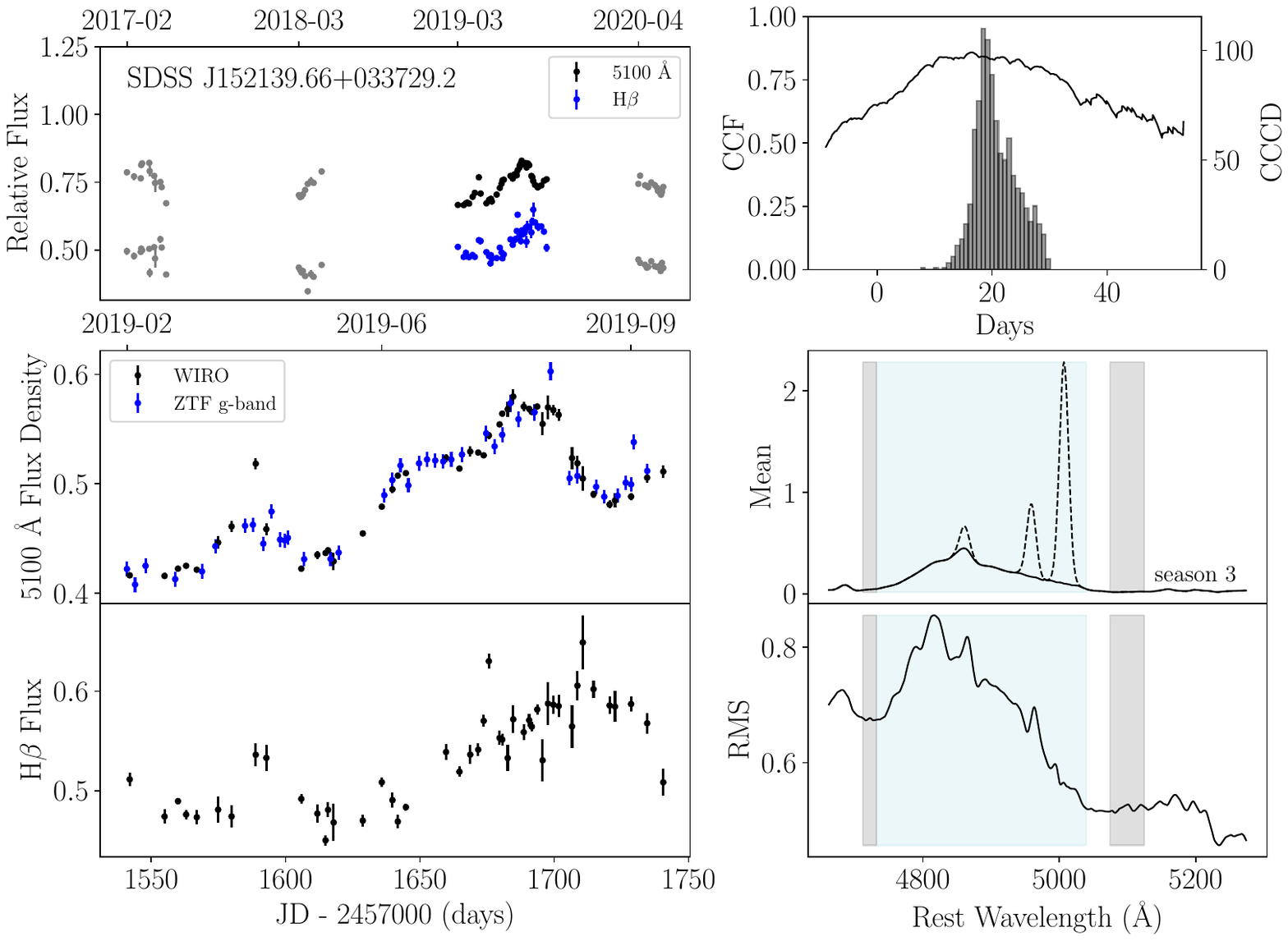}
        \caption{Time-series analysis of SDSS J152139.66+033729.2. The meanings of the panels are the same as Figure~\ref{figure:1es_lc}.}
        \label{figure:sdss1521_lc}
        \end{figure*}

\begin{figure*}[t]
        \includegraphics[width=1\textwidth]{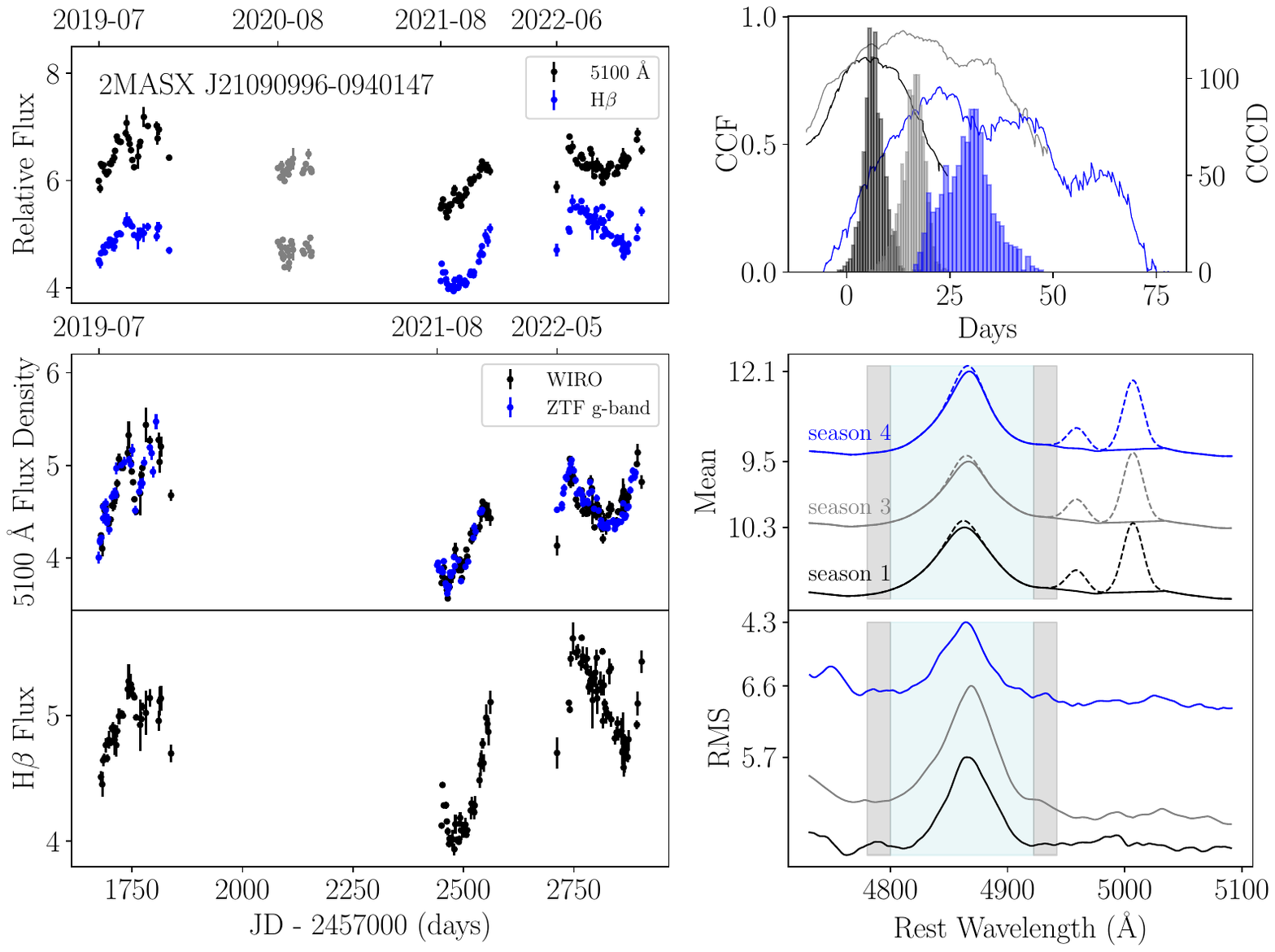}
        \caption{Time-series analysis of 2MASX J21090996-0940147. The meanings of the panels are the same as Figure~\ref{figure:1es_lc}.}
        \label{figure:2masx_lc}
        \end{figure*}

\begin{figure*}[t]
        \includegraphics[width=1\textwidth]{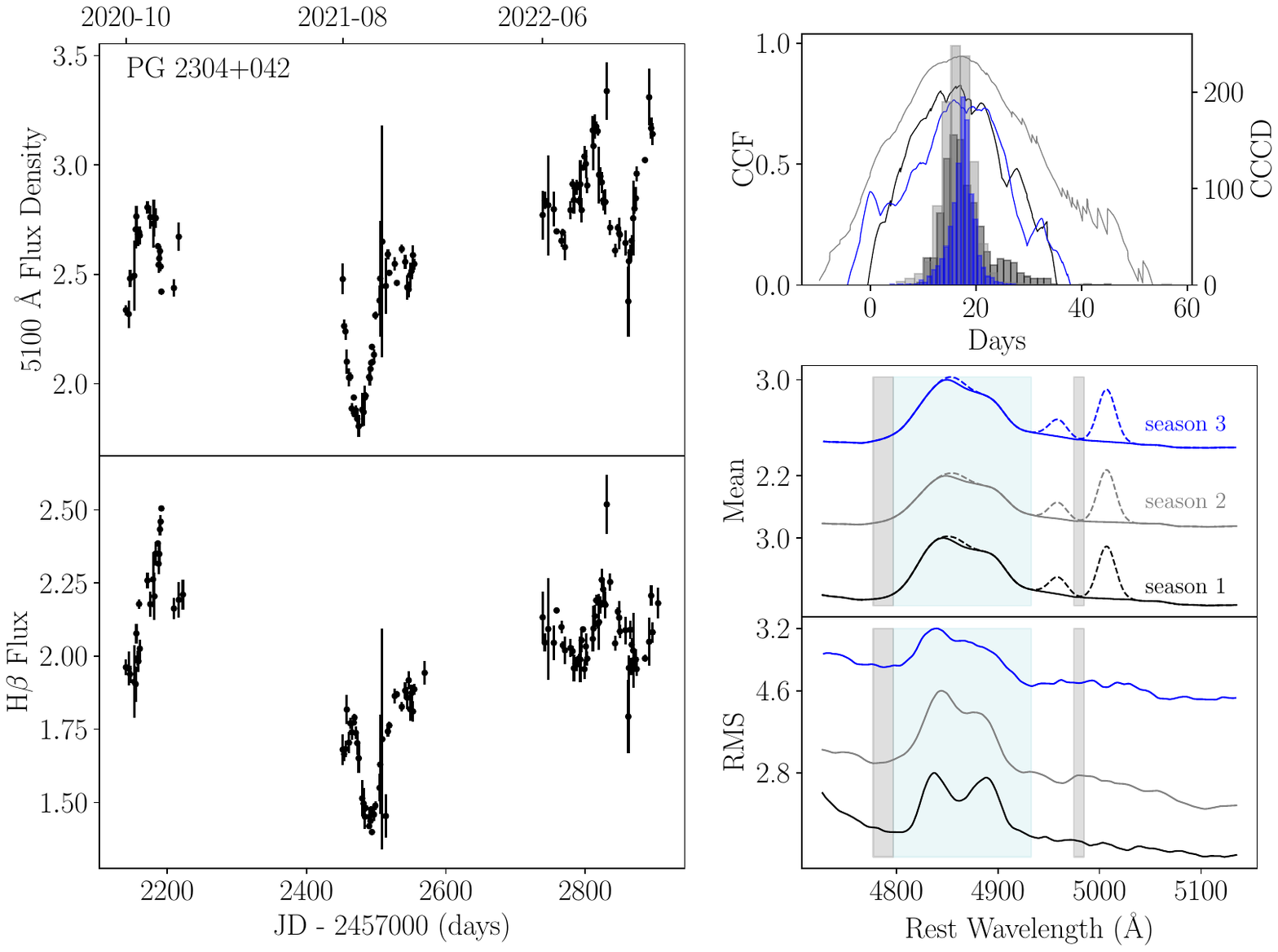}
        \caption{Time-series analysis of PG 2304+042. There are only two left panels as we analyze all WIRO seasons for this object. The meanings of the panels are the same as Figure~\ref{figure:1es_lc}.}
        \label{figure:pg2304_lc}
        \end{figure*}

\begin{figure*}[t]
        \includegraphics[width=1\textwidth]{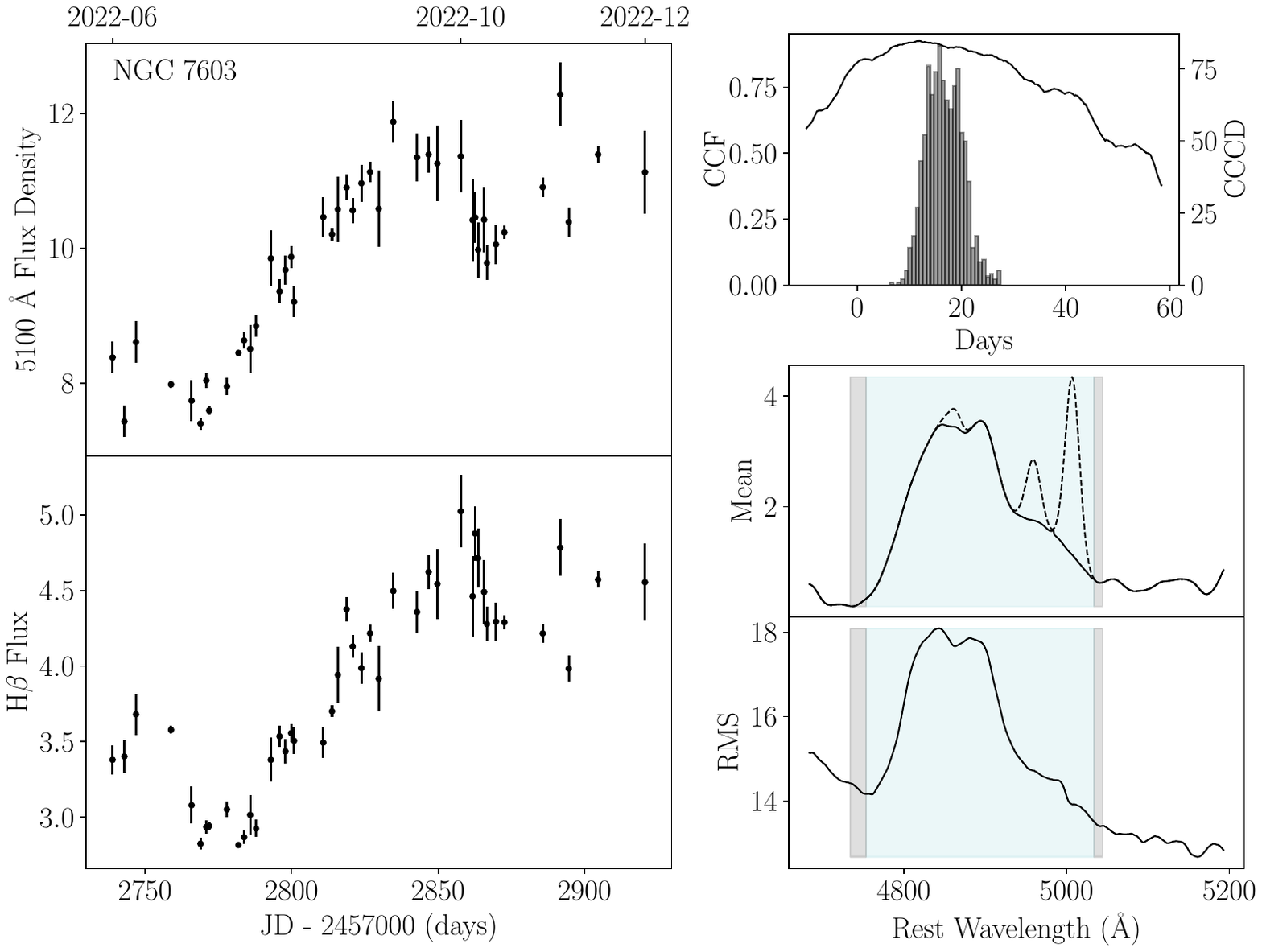}
        \caption{Time-series analysis of NGC 7603. There are only two left panels as we analyze all WIRO seasons for this object. The meanings of the panels are the same as Figure~\ref{figure:1es_lc}.}
        \label{figure:ngc7603_lc}
        \end{figure*}

See Table~\ref{tab:lc_stats} for light curve statistics, including the flux density and flux variation amplitude and uncertainty defined by \citet{Rodriguez-Pascal197} as   
\begin{equation}
    F_{var} = \frac{(\sigma^2 - \Delta^2)^{1/2}}{\langle F \rangle}
\end{equation}

\begin{equation}
    \sigma_{var} = \frac{1}{(2N)^{1/2}F_{var}} \frac{\sigma^2}{\langle F \rangle^2}
\end{equation}
respectively. See WIRO light curve data in Table~\ref{tab:wiro_lc}, and object light curves in Figures~\ref{figure:1es_lc}-\ref{figure:ngc7603_lc}.

\begin{deluxetable*}{lcccccccc}
\tablecaption{Light Curve Statistics\label{tab:lc_stats}}
\tablewidth{0pt}
\tablehead{
\colhead{Object} & \colhead{Season} & \colhead{Duration} & \colhead{N$_{spec}$} & \colhead{Average Cadence} & \colhead{$\bar{\rm{F}}_{5100}$} & \colhead{F$_{5100}$ F$_{var}$} & \colhead{$\bar{\rm{F}}_{\rm{H}\beta}$} & \colhead{H$\beta$ F$_{var}$}\\
\colhead{} & \colhead{} & \colhead{} & \colhead{} & \colhead{(Days)} & \colhead{} & \colhead{(\%)} & \colhead{} & \colhead{(\%)}\\
\colhead{(1)} & \colhead{(2)} & \colhead{(3)} & \colhead{(4)} & \colhead{(5)} & \colhead{(6)} & \colhead{(7)} & \colhead{(8)} & \colhead{(9)}
}

\startdata
1ES 0206+522 & 1 & 2018.08-2019.04 & 68 & 3.36 & 2.31 $\pm$ {0.45} & 0.19 $\pm$ {0.01} & 2.47 $\pm$ {0.13} & 0.03 $\pm$ {0.01} \\
& 2 & 2019.07-2020.04 & 67 & 4.09 & 1.88 $\pm$ {0.22} & 0.12 $\pm$ {0.01} & 2.01 $\pm$ {0.16} & 0.06 $\pm$ {0.01} \\
& 5 & 2022.08-2023.02 & 36 & 5.18 & 0.93 $\pm$ {0.12} & 0.13 $\pm$ {0.01} & 1.59 $\pm$ {0.17} & 0.10 $\pm$ {0.01} \\
Mrk 1040 & 1 & 2018.08-2019.03 & 61 & 3.32 & 3.76 $\pm$ {0.16} & 0.04 $\pm$ {0.004} & 2.23 $\pm$ {0.12} & 0.05 $\pm$ {0.01} \\
& 3 & 2022.07-2023.03 & 39 & 6.15 & 3.91 $\pm$ {0.34} & 0.08 $\pm$ {0.01} & 2.21 $\pm$ {0.22} & 0.10 $\pm$ {0.01}\\
Mrk 618 & 1 & 2019.09-2020.03 & 19 & 8.56 & 4.73 $\pm$ {0.28} & 0.06 $\pm$ {0.01} & 2.97 $\pm$ {0.13} & 0.04 $\pm$ {0.01}\\
& 2 & 2020.09-2021.03 & 48 & 4.20 & 4.51 $\pm$ {0.50} & 0.11 $\pm$ {0.01} & 3.30 $\pm$ {0.19} & 0.05 $\pm$ {0.01} \\
& 3 & 2021.08-2022.02 & 49 & 3.52 & 3.55 $\pm$ {0.18} & 0.05 $\pm$ {0.01} & 2.90 $\pm$ {0.96} & 0.02 $\pm$ {0.01} \\
& 4 & 2022.08-2023.02 & 38 & 4.94 & 3.31 $\pm$ {0.28} & 0.08 $\pm$ {0.01} & 2.89 $\pm$ {0.26} & 0.09 $\pm$ {0.01} \\
MCG -02-14-009 & 3 & 2020.09-2021.03 & 25 & 7.19 & 2.67 $\pm$ {0.18} & 0.07 $\pm$ {0.003} & 2.15 $\pm$ {0.14} & 0.05 $\pm$ {0.01} \\
& 4 & 2021.09-2022.03 & 40 & 4.87 & 2.45 $\pm$ {0.22} & 0.09 $\pm$ {0.01} & 2.14 $\pm$ {0.14} & 0.05 $\pm$ {0.01} \\
& 5 & 2022.09-2023.03 & 30 & 6.29 & 2.93 $\pm$ {0.20} & 0.07 $\pm$ {0.004} & 2.24 $\pm$ {0.21} & 0.09 $\pm$ {0.01} \\
IRAS 05589+2828 & 1 & 2018.09-2019.03 & 56 & 3.32 & 3.62 $\pm$ {0.15} & 0.04 $\pm$ {0.003} & 8.00 $\pm$ {0.32} & 0.04 $\pm$ {0.004} \\
& 2 & 2019.08-2020.05 & 57 & 4.75 & 3.86 $\pm$ {0.20} & 0.05 $\pm$ {0.003} & 7.89 $\pm$ {0.36} & 0.04 $\pm$ {0.004} \\
& 3 & 2020.08-2021.04 & 73 & 3.37 & 2.67 $\pm$ {0.16} & 0.06 $\pm$ {0.003} & 5.08 $\pm$ {0.33} & 0.06 $\pm$ {0.01} \\
& 4 & 2021.08-2022.04 & 74 & 3.33 & 3.70 $\pm$ {0.43} & 0.12 $\pm$ {0.01} & 6.64 $\pm$ {0.92} & 0.14 $\pm$ {0.01} \\
& 5 & 2022.08-2023.05 & 52 & 5.19 & 3.44 $\pm$ {0.23} & 0.06 $\pm$ {0.01} & 7.26 $\pm$ {0.52} & 0.07 $\pm$ {0.01} \\
Mrk 715 & 4 & 2020.02-2020.05 & 10 & 11.39 & 1.57 $\pm$ {0.91} & 0.06 $\pm$ {0.01} & 0.58 $\pm$ {0.06} & 0.09 $\pm$ {0.03} \\
& 5 & 2020.11-2021.05 & 20 & 9.19 & 1.35 $\pm$ {0.13} & 0.09 $\pm$ {0.01} & 0.46 $\pm$ {0.09} & 0.18 $\pm$ {0.03} \\
SBS 1136+594 & 1 & 2022.01-2022.08 & 41 & 5.02 & 1.52 $\pm$ {0.11} & 0.07 $\pm$ {0.01} & 1.60 $\pm$ {0.12} & 0.07 $\pm$ {0.01} \\
& 2 & 2022.10-2023.05 & 40 & 5.62 & 1.66 $\pm$ {0.16} & 0.09 $\pm$ {0.01} & 1.48 $\pm$ {0.08} & 0.05 $\pm$ {0.01} \\
VIII Zw 233 & 2 & 2017.11-2018.05 & 35 & 4.88 & 0.64 $\pm$ {0.06} & 0.09 $\pm$ {0.01} & 0.22 $\pm$ {0.03} & 0.13 $\pm$ {0.02} \\
& 3 & 2019.01-2019.07 & 28 & 6.45 & 0.62 $\pm$ {0.02} & 0.03 $\pm$ {0.002} & 0.19 $\pm$ {0.01} & 0.05 $\pm$ {0.01} \\
Mrk 813 & 4 & 2021.09-2022.09 & 29 & 8.13 & 1.85 $\pm$ {0.14} & 0.07 $\pm$ {0.01} & 1.45 $\pm$ {0.08} & 0.05 $\pm$ {0.01} \\
SDSS J145307.92+255433.0 & 1 & 2022.02-2022.09 & 18 & 11.81 & 1.07 $\pm$ {0.12} & 0.11 $\pm$ {0.01} & 0.61 $\pm$ {0.05} & 0.07 $\pm$ {0.02} \\
SDSS J152139.66+033729.2 & 3 & 2019.02-2019.09 & 46 & 4.32 & 0.50 $\pm$ {0.05} & 0.10 $\pm$ {0.08} & 0.54 $\pm$ {0.05} & 0.09 $\pm$ {0.01} \\
2MASX J21090996-0940147 & 1 & 2019.07-2019.12 & 33 & 4.81 & 4.75 $\pm$ {0.34} & 0.07 $\pm$ {0.01} & 4.93 $\pm$ {0.21} & 0.04 $\pm$ {0.01} \\
& 3 & 2021.08-2021.12 & 37 & 2.99 & 4.01 $\pm$ {0.28} & 0.07 $\pm$ {0.01} & 4.28 $\pm$ {0.32} & 0.07 $\pm$ {0.01} \\
& 4 & 2022.05-2022.11 & 56 & 3.42 & 4.59 $\pm$ {0.21} & 0.04 $\pm$ {0.003} & 5.12 $\pm$ {0.28} & 0.05 $\pm$ {0.01} \\
PG 2304+042 & 1 & 2020.10-2021.01 & 24 & 3.41 & 2.61 $\pm$ {0.14} & 0.05 $\pm$ {0.01} & 2.17 $\pm$ {0.18} & 0.08 $\pm$ {0.01} \\
& 2 & 2021.08-2021.12 & 41 & 2.87 & 2.23 $\pm$ {0.27} & 0.11 $\pm$ {0.02} & 1.67 $\pm$ {0.17} & 0.09 $\pm$ {0.01} \\
& 3 & 2022.06-2022.11 & 52 & 3.19 & 2.86 $\pm$ {0.20} & 0.07 $\pm$ {0.01} & 2.08 $\pm$ {0.11} & 0.05 $\pm$ {0.01} \\
NGC 7603 & 1 & 2022.06-2023.01 & 43 & 4.23 & 9.85 $\pm$ {1.31} & 0.13 $\pm$ {0.02} & 4.71 $\pm$ {0.73} & 0.15 $\pm$ {0.02} 
\enddata

\tablecomments{Column 1 is the object name. Column 2 is the season (for a graphical sense of the meaning behind "season" see Figure~\ref{figure:1es_lc}). Columns 3-5 are the duration, number of epochs and average cadence of the spectroscopy in days. Columns 6 \& 7 are the mean flux density in units of 10$^{-15}$ erg s$^{-1}$ cm$^{-2}$ \r{A}$^{-1}$ and the flux density variance of the continuum light curve. Columns 8 \& 9 are the mean flux in  units of 10$^{-13}$ erg s$^{-1}$ cm$^{-2}$ and the flux variance of the H$\beta$ light curve. The uncertainty on the mean flux densities and fluxes is the standard deviation of the light curves.}
\end{deluxetable*}

\begin{deluxetable*}{lcccccc}
\tablecaption{Light Curves\label{tab:wiro_lc}}
\tablewidth{0pt}
\tablehead{
\colhead{Object} & \colhead{Season} & \colhead{JD-2457000} & \colhead{Telescope} & \colhead{Data} & \colhead{Flux (density)} & \colhead{Flux (density) uncertainty} \\
\colhead{(1)} & \colhead{(2)} & \colhead{(3)} & \colhead{(4)} & \colhead{(5)} & \colhead{(6)} & \colhead{(7)} 
}
\startdata
1ES 0206+522 & 1 & 1340.96830 & ZTF & Cont & 2.253 & 0.029 \\
1ES 0206+522 & 1 & 1342.95124 & ZTF & Cont & 2.243 & 0.029 \\ 
1ES 0206+522 & 1 & 1354.90833 & WIRO & Cont & 2.187 & 0.070 \\
1ES 0206+522 & 1 & 1354.90833 & WIRO & H$\beta$ & 1.713 & 0.055 \\
1ES 0206+522 & 1 & 1358.96782 & WIRO & H$\beta$ & 1.747 & 0.107 \\
1ES 0206+522 & 1 & 1359.87058 & WIRO & H$\beta$ & 1.733 & 0.044 
\enddata

\tablecomments{The units for the continuum flux densities and uncertainties and H$\beta$ fluxes and uncertainties are 10$^{-15}$ erg s$^{-1}$ cm$^{-2}$ \r{A}$^{-1}$ and 10$^{-13}$ erg s$^{-1}$ cm$^{-2}$ respectively.\\
(This table is available in its entirety in machine-readable form).}

\end{deluxetable*}

\section{Analysis} \label{sec:analysis}
\subsection{Time-Series Analysis} \label{subsec:time_lags}

We measured the delay between 5100 $\textrm{\r{A}}$ continuum variations and H$\beta$ line response using the interpolated cross-correlation function \citep[ICCF;][]{GaskellSparke1986,GaskellPeterson1987,WhitePeterson1994}. The ICCF measures the strength of the correlation between the two light curves at a range of lags. It does this twice, alternating which light curve is interpolated. The output cross correlation function (CCF) is the average of the two. We can determine the time delay from the ICCF in two ways: the lag corresponding to the the peak correlation coefficient, r, or the centroid of some range of correlation coefficients (generally $>$ 0.8 r$_{\textrm{max}}$ where r$_{\textrm{max}}$ is the maximum correlation coefficient). If the BLR has an extended geometry, the CCF peak is biased towards the inner radius while the centroid is a less biased, luminosity-weighted measurement \citep{KoratkarGaskell1991,Perez1992}. The centroid is also more consistent with the virialized motion of the BLR gas \citep{WhitePeterson1994}. 

To investigate time-lag uncertainties we used the “flux randomization/random subset sampling” (FR/RSS) method \citep{Peterson1998_2}. The FR/RSS method accounts for measurement and sampling uncertainties. We generated mock light curves by perturbing the measured flux density or flux values according to their uncertainties and randomly re-sampling our light curves with repetition. The FR accounts for measurement uncertainties and the RSS accounts for uncertainty due to inclusion/exclusion of any one point. We used the FR/RSS method in a Monte Carlo fashion, running it 1000 times to build up cross-correlation centroid (peak) distribution (CCCD or CCPD). We take the median and 68\% confidence interval of the CCCD as our lag and uncertainty. We report the median of the CCPD with uncertainties for completeness. 

We chose the lag search range for the CCF for each object based on visual inspection of the CCFs and CCCDs, choosing the upper limit that allows the CCF to reach a minimum. Our search windows all begin at $-$10 days and increase up to 105 days. 

\citet{Welsh1999} showed that the CCF measurements can be strongly affected when a light curve has a varying mean or variance across the time sampled. In other words, the presence of long-term secular trends in a light curve will bias the time-lag measurement toward smaller lags \citep[see also][]{Perez1992}. The process of fitting and subtracting a low-order polynomial to the light curves is called detrending and has been explored in various RM studies \citep[e.g.,][]{Grier2008, Peterson2014}. We note that detrending might be removing long-term variation that is part of the reverberation response. We therefore proceed with caution and only detrend one light curve (see the discussion in Section~\ref{subsec:1es}) where there is a clear slope in the continuum light curve while the H$\beta$ light curve stays mostly flat. Our detrending process consists of fitting a line to the light curve using non-linear least squares minimization and then subtracting that line from the light curve. We report time lags and lag search ranges in Table~\ref{tab:widthlags}.

\subsection{Line-Width Measurements} \label{subsec:line_width}
The H$\beta$ line is broad and can be contaminated by the narrow [O III] requiring a narrow-line subtraction in the mean spectra to accurately measure the line width. It is not necessary to subtract narrow lines from the rms spectra, as the narrow lines do not vary on RM time scales \citep{Peterson2013}. We followed the procedure outlined in MAHA I, fitting [O III] $\lambda$5007 as a single Gaussian and using the best-fit parameters as a template to fit [O III] $\lambda$4959. We assumed that the narrow component of H$\beta$ has the same profile as [O III] and used our template to fit. We fixed the separation between [O III] $\lambda$5007, [O III] $\lambda$4959, and H$\beta$ to their laboratory separations. The [O III] $\lambda$5007/[O III] $\lambda$4959 and [O III] $\lambda$5007/H$\beta$ flux ratios measured from our line fits are close to typical AGN values, 2.98 and 10, respectively \citep{Kewley2006,SternLaor2013}, indicating the fitting procedure outlined here works well. See Table~\ref{tab:flux_ratios} for the narrow-line flux ratios for each season of each object. 

\begin{deluxetable*}{lccc}
\tablecaption{Narrow-line Flux Ratios\label{tab:flux_ratios}}
\tablewidth{0pt}
\tablehead{
\colhead{Object} & \colhead{Season} & \colhead{[O III]$\lambda$5007/[O III]$\lambda$4959} & \colhead{[O III]$\lambda$5007/H$\beta$} \\
\colhead{(1)} & \colhead{(2)} & \colhead{(3)} & \colhead{(4)}
}
\startdata
1ES 0206+522 & 1 & 3.24 & 10.37 \\
& 2 & 3.22 & 10.20 \\
& 5 & 3.11 & 9.99 \\
Mrk 1040 & 1 & 3.25 & 10.22 \\
& 3 & 3.38 & 10.99 \\
Mrk 618 & 1 & 3.52 & 10.98 \\
& 2 & 3.52 & 10.98 \\
& 3 & 3.55 & 10.98 \\
& 4 & 3.42 & 10.51 \\
MCG -02-14-009 & 3 & 3.29 & 10.99 \\
& 4 & 3.52 & 10.99 \\
& 5 & 3.68 & 10.99 \\
IRAS 05589+2828 & 1 & 3.55 & 9.26 \\
& 2 & 3.40 & 9.26 \\
& 3 & 3.31 & 9.26 \\
& 4 & 3.41 & 9.27 \\
& 5 & 3.36 & 9.15 \\
Mrk 715 & 4 & 2.89 & 9.03 \\
& 5 & 2.87 & 8.96 \\
SBS 1136+594 & 1 & 3.08 & 10.00 \\
& 2 & 3.20 & 9.96 \\
VIII Zw 233 & 2 & 3.06 & 10.99 \\
& 3 & 3.04 & 10.99 \\
Mrk 813 & 4 & 3.82 & 9.45 \\
SDSS J145307.92+255433.0.0 & 1 & 2.97 & 9.99 \\
SDSS J152139.66+033729.2 & 3 & 3.05 & 10.00 \\
2MASX J21090996-0940147 & 1 & 3.31 & 9.98 \\
& 3 & 3.28 & 9.98 \\
& 4 & 3.41 & 9.98 \\
PG 2304+042 & 1 & 3.11 & 10.47 \\
& 2 & 3.23 & 10.46 \\
& 3 & 3.08 & 10.48 \\
NGC 7603 & 1 & 2.91 & 9.94
\enddata

\tablecomments{Column 1 contains the object name. Column 2 lists the season. Column 3 lists the flux ratio of [O III]$\lambda$5007/[O III]$\lambda$4959. Column 4 lists the flux ratio of [O III]$\lambda$5007/H$\beta$.}
\end{deluxetable*}

We used two methods to measure line widths: the FWHM and the line dispersion, $\sigma_{\rm{H}\beta}$. As many of our objects have significantly asymmetric or double-peaked H$\beta$ profiles, we followed the procedure in \citet{Peterson2004} to measure the FWHM of double-peaked profiles, by considering the blue peak and the red peak separately. The second moment of the line profile is the mean square dispersion of the line

    \begin{equation}
        \sigma_{\rm line}^2(\lambda) = \langle \lambda^2 \rangle - \lambda_0^2 = \left[ \int \lambda^2 P(\lambda)d\lambda / \int P\lambda d\lambda \right] - \lambda_0^2.
    \end{equation}
We take the square root of the second moment as $\sigma_{\rm{H}\beta}$.

To evaluate the uncertainty in our line widths we follow the procedure established in MAHA I of using the bootstrap method. We resampled the mean/rms spectrum for each object, using N randomly selected points with repetition, out of the N points in the spectrum, where N in both cases is the number of points in the spectrum. We then measured the FWHM and $\sigma_{\rm{H}\beta}$ from the resampled spectrum. We did this 500 times and took the median and standard deviation of the resulting distribution as our line-width measurement and uncertainty. 

Finally, we subtracted instrumental line broadening from our line widths. MAHA I estimates the instrumental line spread function by comparing the [O III] widths to the intrinsic widths reported in \citet{Whittle1992}. Following MAHA I, we adopted 925 km s$^{-1}$ as the average line-spread function and subtracted it in quadrature from all line widths. This correction introduces a small uncertainty into the line width measurements. We report line widths in Table ~\ref{tab:widthlags}

\subsection{Black Hole Masses} \label{subsec:bh_mass}
The mass of the supermassive black hole fueling the AGN can be estimated assuming virialized motion \citep{PetersonWandel1999,Wandel1999} in the BLR using 
    \begin{equation}
       M_{\bullet} = f_{\rm BLR} \frac{R_{\rm BLR} \Delta V^2}{G} = f_{\rm BLR} M_{\rm VP}
    \end{equation}
where R$_{\rm{BLR}}$ = c$\tau_{\rm{BLR}}$ and $\Delta$V$^2$ can be measured from the width of the H$\beta$ line and f$_{\rm{BLR}}$ is a calibration factor that accounts for the unknown geometry of the BLR. It is likely that each object has an individual f$_{\rm{BLR}}$, but dynamical modeling is needed to find the best fit BLR geometry for a given object and calculate its individual f$_{\rm{BLR}}$ \citep[e.g.,][]{Pancoast2014, Li2013, Li2018}. As none of our sample targets have individual f$_{\rm{BLR}}$ values in the literature, we use an average f$_{\rm{BLR}}$ that is calibrated to bring active galaxy masses in line with the quiescent masses according to the M$_{\bullet}$-$\sigma_*$ or the M$_{\bullet}$-M$_*$ quiescent galaxy relationships \citep[e.g.,][]{Onken2004,Woo2010,Woo2015,Graham2011,Park2012,Grier2013,HoKim2014,Batiste2017}. This average f$_{\rm{BLR}}$ is appropriate for a population of AGNs, but introduces an uncertainty of a factor of a few in individual objects. Following previous MAHA papers, we adopted the empirically calibrated average value from \citet{Woo2015} of f$_{\rm{BLR}}$ 1.12 (4.47) when using FWHM  ($\sigma_{\rm{H}\beta}$) from the rms spectra. We do not include the uncertainty on f$_{\rm{BLR}}$ in our SMBH mass uncertainty calculations.

Some care must be taken in the choice of lag and line-width measurement. As discussed in section \ref{subsec:time_lags}, we chose the median of the CCCD as our lag measurement. Regarding the line measurement, there are two choices to be made: mean or rms spectrum and FWHM or $\sigma_{\rm{H}\beta}$. The advantage of the rms spectrum is that it only displays the variable part of the spectrum, meaning that narrow lines and other non-varying components drop out and we do not have to worry about them contaminating the broad-line width measurement. This removes the need to do spectral fitting which can be ambiguous. Also, in accordance with the assumption of virialized motion in the BLR, we measured the lag of the BLR gas that is responding to continuum flux variations and the width of the rms spectrum measures the velocity of this same gas. The rms spectrum can be noisy however, and it can sometimes be difficult to measure the line width. It also can be contaminated by other variable broad lines: He II $\lambda$4686 and Fe II in the case of H$\beta$. Also, the rms requires multiple epochs of spectra to construct, rendering it impossible for use in single-epoch mass estimations. The mean spectrum is generally clean and easily replaced with a single-epoch spectrum. It does require narrow-line fitting and subtraction which can be difficult if the red wing of H$\beta$ blends with [O III] $\lambda\lambda$4959 5007. For these reasons we measured line widths from both the mean and rms spectra, though we prefer the rms spectra because it is intuitive to measure velocities from the variable part of the line.

Regarding FWHM vs $\sigma_{\rm{H}\beta}$, it is important to note they are not necessarily interchangeable and have their own strengths and weaknesses \citep{Peterson2004, Wang2020}. In practice, $\sigma_{\rm{H}\beta}$ better follows the virial assumption, though it is sensitive to the line wings which can be blended with other features. FWHM on the other hand is more easily and precisely measured \citep[though that is not true for the rms spectra which can be noisy][]{Peterson2004}. See \citet{DallaBonta2020} for a recent summary of arguments in favor of $\sigma_{\rm{H}\beta}$. 

Though we report both $\sigma_{\rm{H}\beta}$ and FWHM line widths from both the mean and rms spectra, we only report SMBH masses using $\sigma_{\rm{H}\beta}$ and FWHM measured from the rms spectrum. For completeness and to aid comparison with single-epoch black hole mass estimations, we also report simple virial products (assuming f$_{\rm{H}\beta}$ = 1) measured using the FWHM from the mean spectrum.
We report our masses and virial products in Table~\ref{tab:vp_mass}. We indicate our preferred SMBH mass values for each object with a ``$\checkmark$'' in the last column of Table~\ref{tab:vp_mass}. We prefer SMBH masses that are calculated using $\sigma_{\rm{H}\beta}$ in the rms spectra and have the smallest measurement uncertainties. 

All our reported masses are individually relatively small and collectively span a small range. This is because we are limited to nearby, typically low-luminosity objects observable at WIRO and with time lags short enough to resolve over an observing campaign of several months. In addition, this paper focuses on MAHA objects that provide a result in a single season of observations. Future MAHA papers will report results from objects that require analyzing several observing seasons together to measure a time lag.

\subsection{Velocity-Resolved Time Lags} \label{subsec:vel_res}
It is possible with high S/N, high-cadence RM data to determine average time lags as a function of velocity across the line profile. Traditional RM recovers the average time lag for the observed season, but does not reveal information about the kinematics or geometry of the BLR. In the absence of intensive forward modeling, or potentially difficult to interpret velocity-delay maps, a reasonable method to estimate basic BLR kinematics is to determine the time lag as a function of velocity. MAHA is a high cadence, long-term RM campaign and while dynamical modeling techniques will be used on MAHA data in the future, it is beyond the scope of the present effort. Instead, we measured velocity-resolved time lags and are able to make broad statements about the kinematics of the BLR for the objects presented here. 

Generally, the signatures of a disk-like, outflow, or inflow BLR are as follows: disk-like structures are indicated by shorter lags in the high velocity gas and longer lags in the low velocity gas; outflow shows ``blue leads red" which means longer lags in the redshifted gas and shorter lags in the blue shifted gas; inflow shows ``red leads blue" which means shorter lags in the redshifted gas and longer lags in the blueshifted gas. We note recent work suggesting that sometimes the outflow signature in velocity-resolved time lags can unexpectedly be ``red leads blue" \citep{Mangham2017}. The BLR kinematics elude an easy description and we advise readers to take our interpretations based on velocity-resolved time lags as possibilities rather than certainties in most cases.

\begin{center}
\begin{figure*}
\gridline{\fig{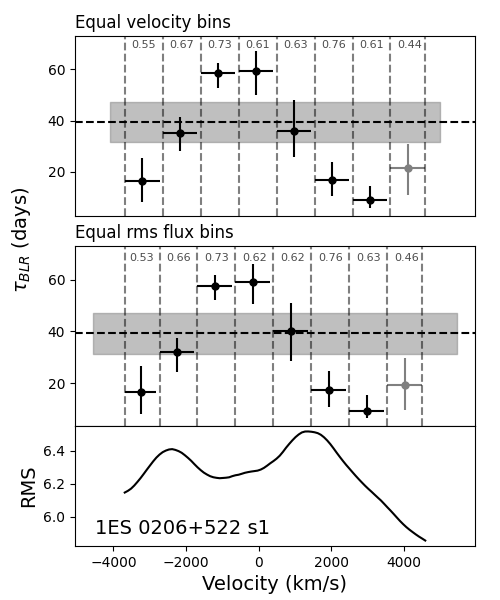}{0.3\textwidth}{}
          \fig{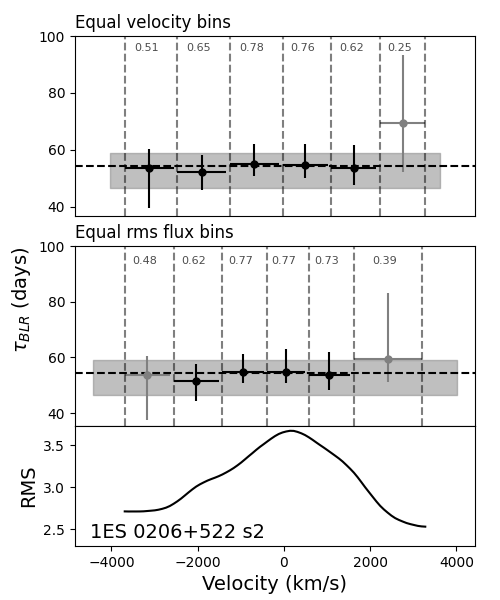}{0.3\textwidth}{}
          \fig{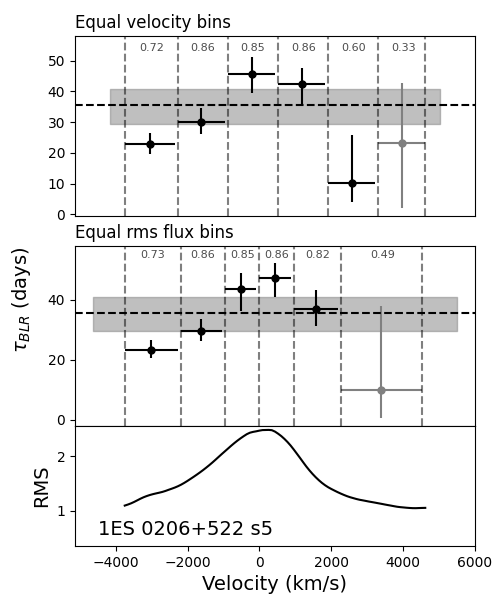}{0.3\textwidth}{}}
\gridline{\fig{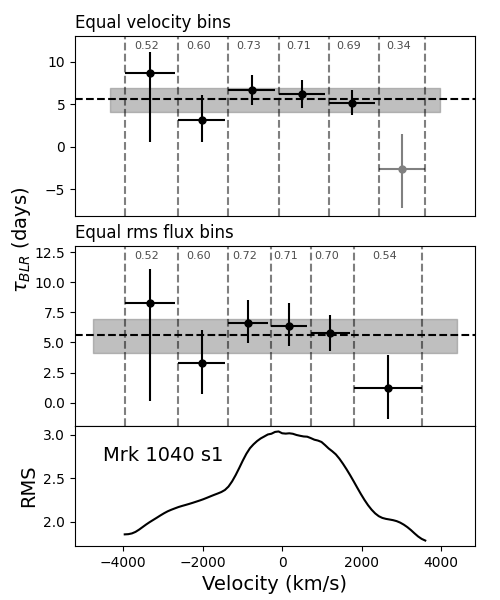}{0.3\textwidth}{}
        \fig{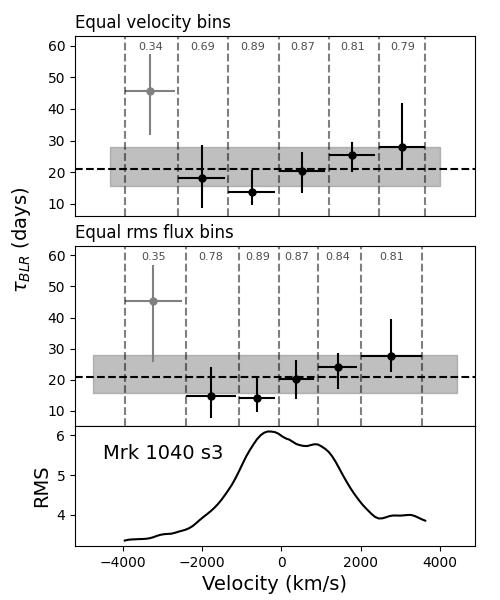}{0.3\textwidth}{}
        \fig{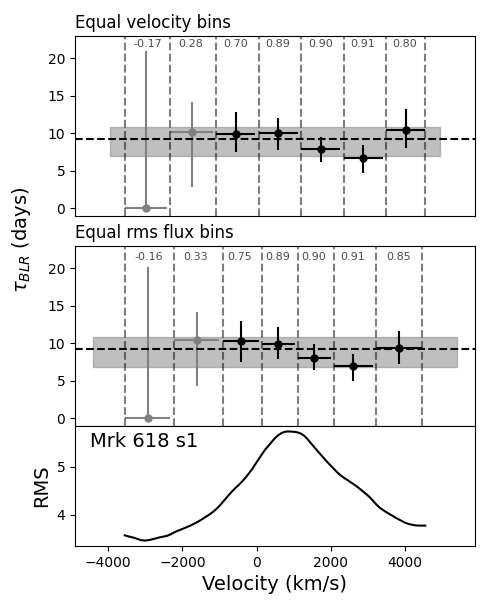}{0.3\textwidth}{}}
\gridline{\fig{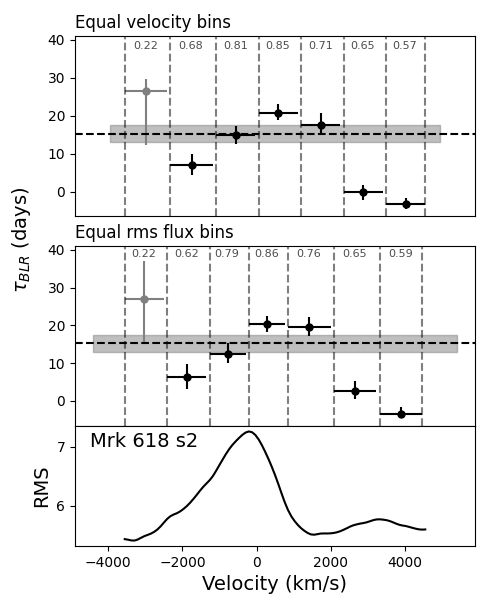}{0.3\textwidth}{}
        \fig{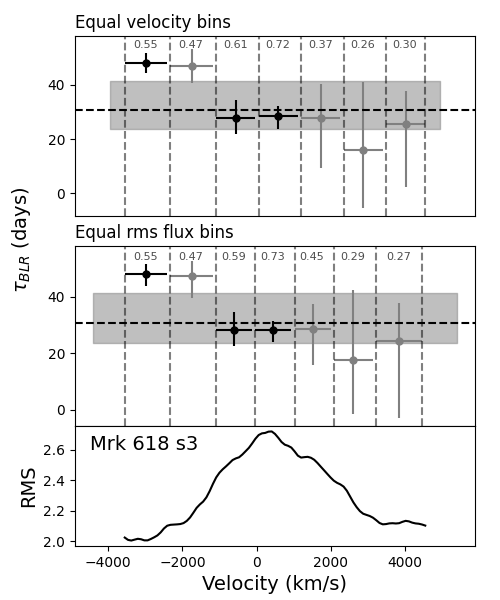}{0.3\textwidth}{}
        \fig{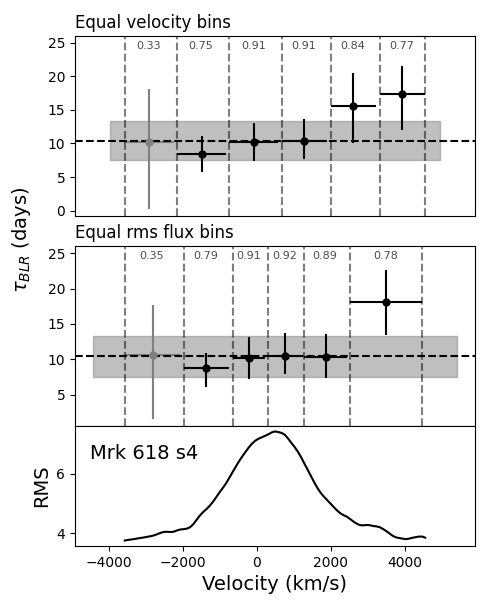}{0.3\textwidth}{}}
\caption{Velocity-resolved lags. The top panel in each subplot shows the lags using bins of equal velocity width. The middle panel shows the lags using bins of equal rms flux density. In both the top and middle panels, the cross-correlation coefficients for each bin are written at the top of the bin. The bottom panel shows the rms spectrum without continuum subtraction in units of 10$^{-16}$ erg s$^{-1}$ cm$^{-2}$ \r{A}$^{-1}$. The bottom panel also shows the object name and season number. Grayed-out points are lag measurements with cross-correlation coefficients less than 0.5.}
\label{figure:v_resolved1}
\end{figure*}
\end{center}

\begin{center}
\begin{figure*}
\gridline{\fig{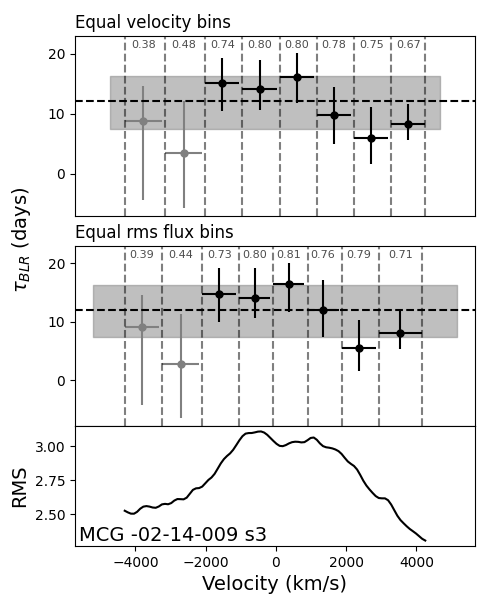}{0.3\textwidth}{}
        \fig{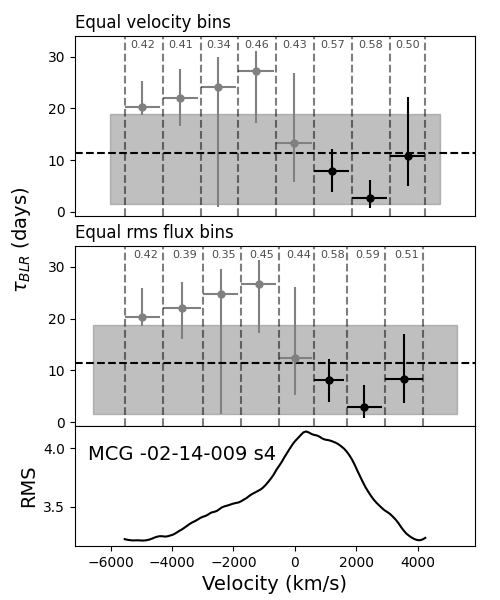}{0.3\textwidth}{}
        \fig{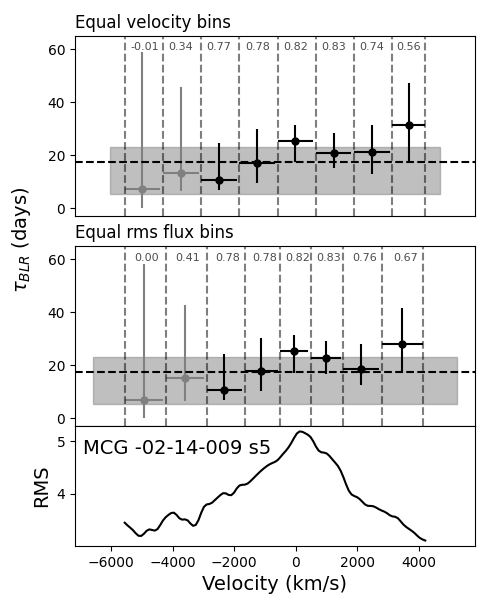}{0.3\textwidth}{}}
\gridline{\fig{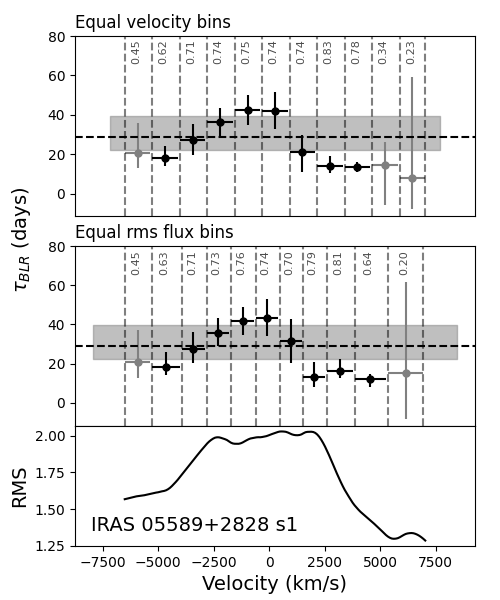}{0.3\textwidth}{}
        \fig{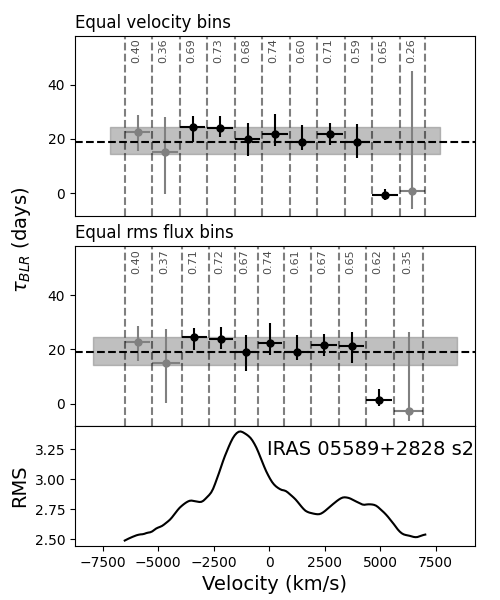}{0.3\textwidth}{}
        \fig{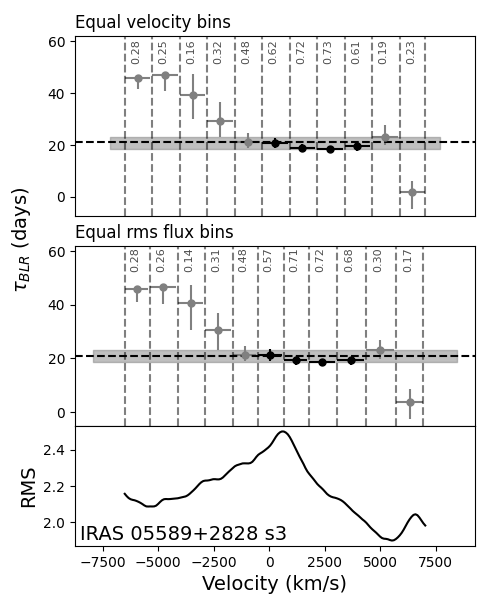}{0.3\textwidth}{}}
\gridline{\fig{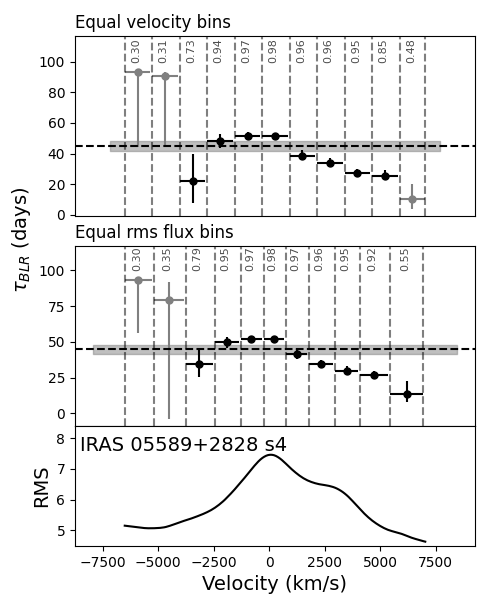}{0.3\textwidth}{}
        \fig{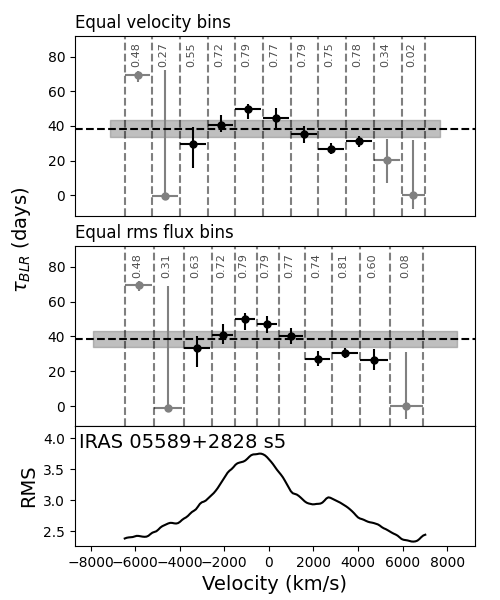}{0.3\textwidth}{}
        \fig{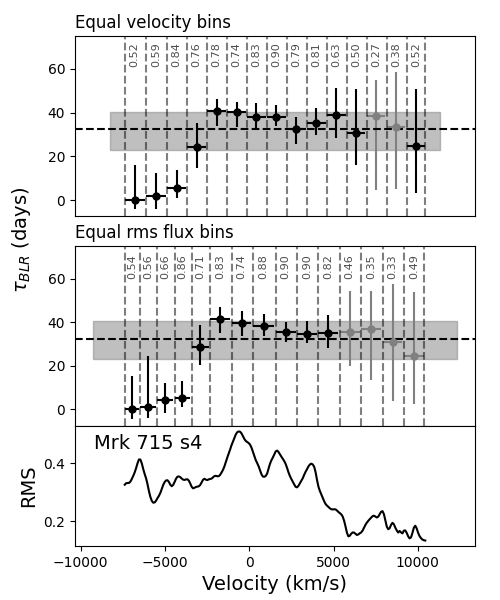}{0.3\textwidth}{}}
\caption{Velocity-resolved lags. The meanings of the panels are the same as in Figure~\ref{figure:v_resolved1}}
\label{figure:v_resolved2}
\end{figure*}
\end{center}

\begin{center}
\begin{figure*}
\gridline{\fig{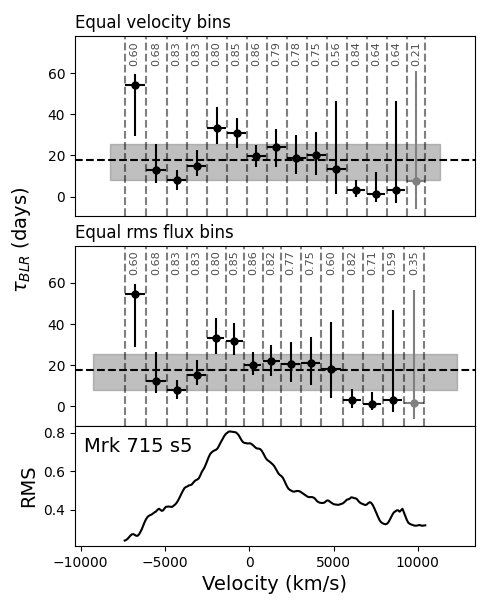}{0.3\textwidth}{}
        \fig{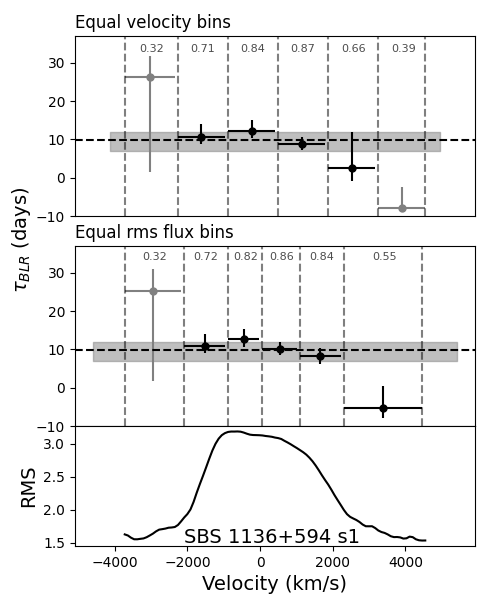}{0.3\textwidth}{}
        \fig{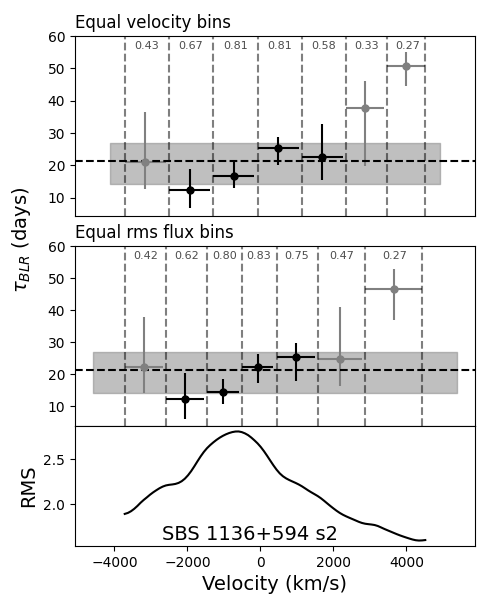}{0.3\textwidth}{}}
\gridline{\fig{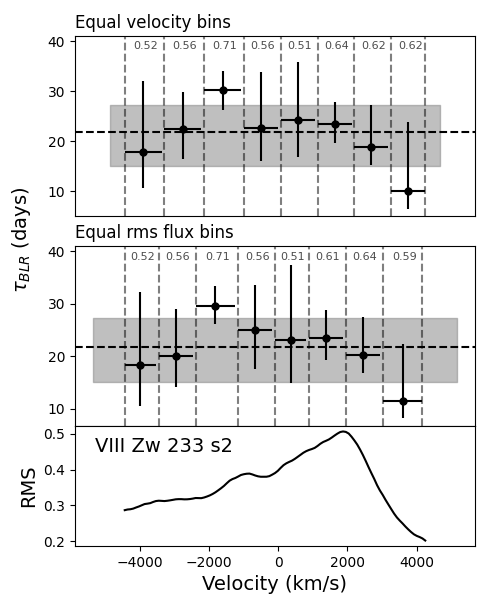}{0.3\textwidth}{}
        \fig{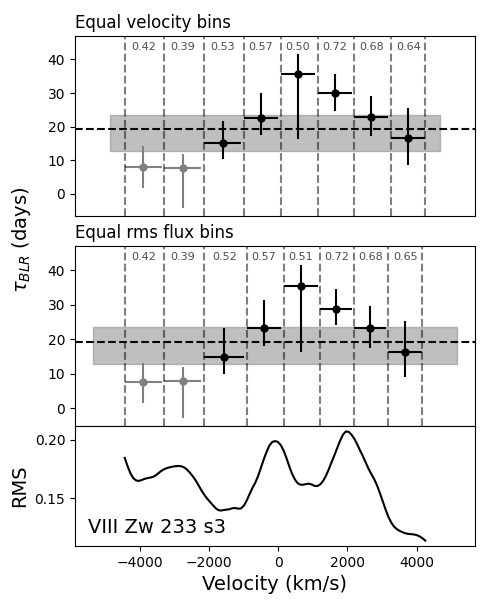}{0.3\textwidth}{}
        \fig{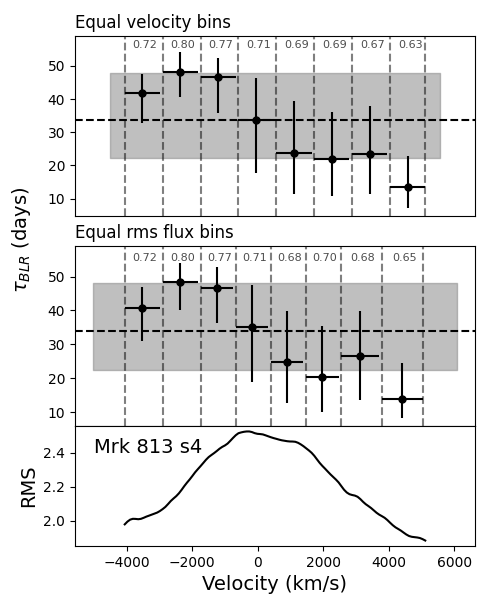}{0.3\textwidth}{}}
\gridline{\fig{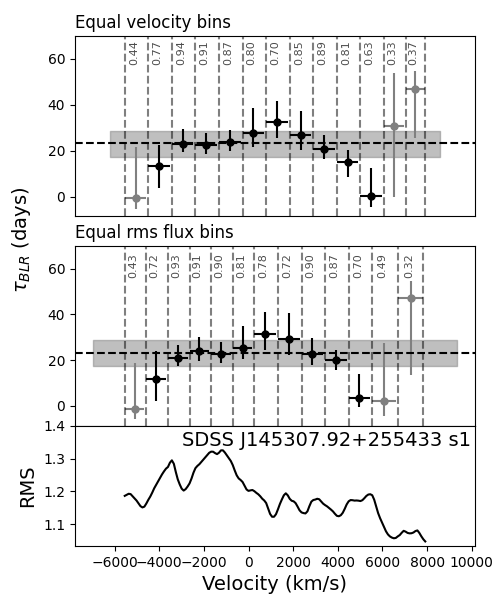}{0.3\textwidth}{}
        \fig{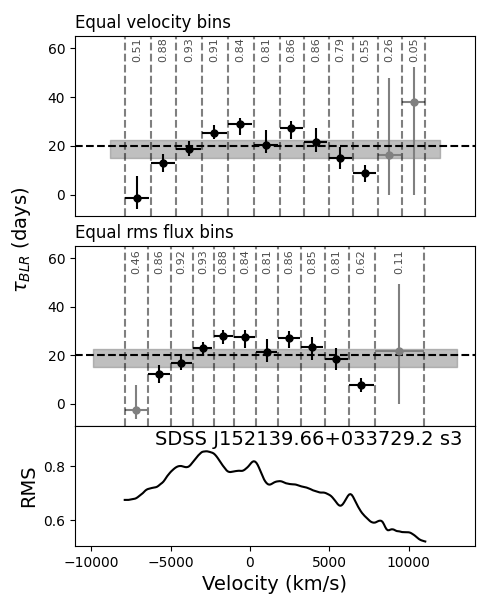}{0.3\textwidth}{}
        \fig{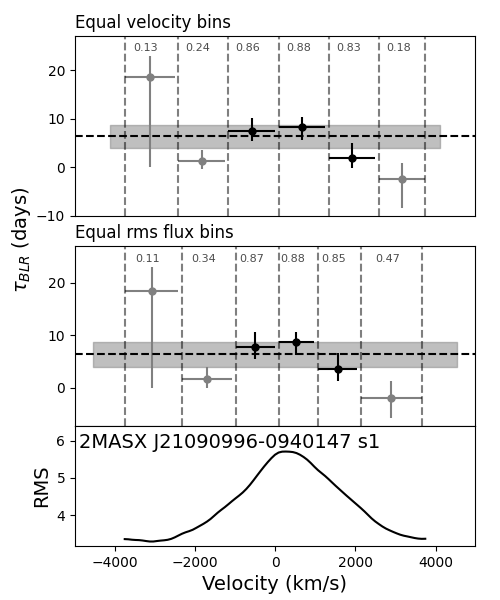}{0.3\textwidth}{}}
\caption{Velocity-resolved lags. The meanings of the panels are the same as in Figure~\ref{figure:v_resolved1}}
\label{figure:v_resolved3}
\end{figure*}
\end{center}

\begin{center}
\begin{figure*}
\gridline{\fig{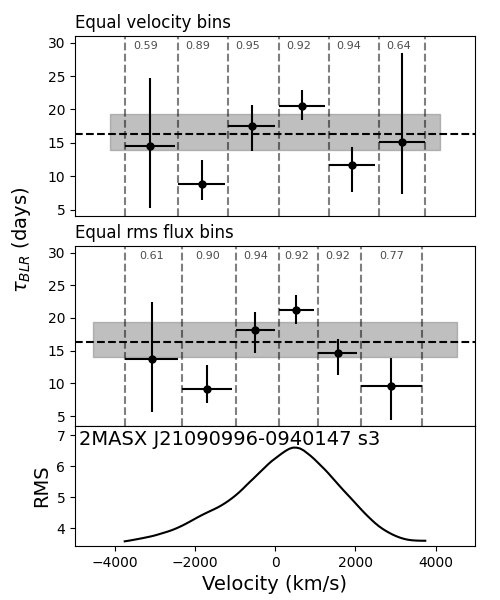}{0.3\textwidth}{}
        \fig{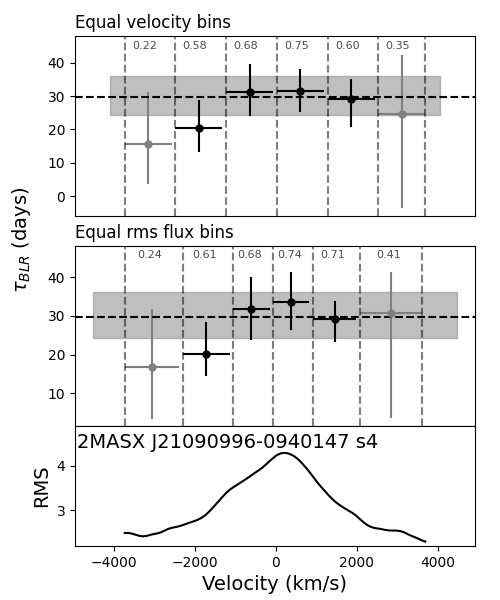}{0.3\textwidth}{}
        \fig{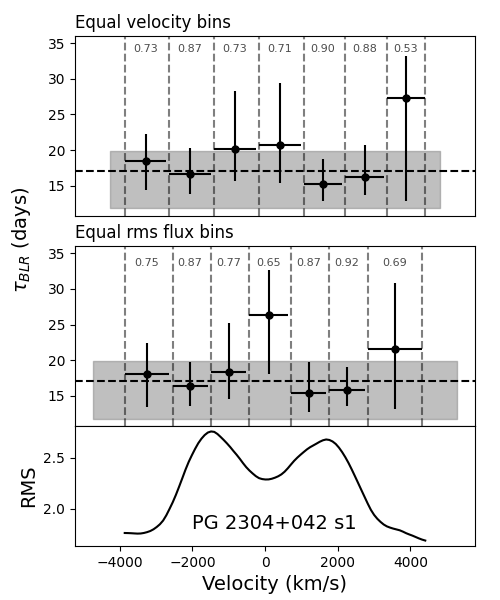}{0.3\textwidth}{}}
\gridline{\fig{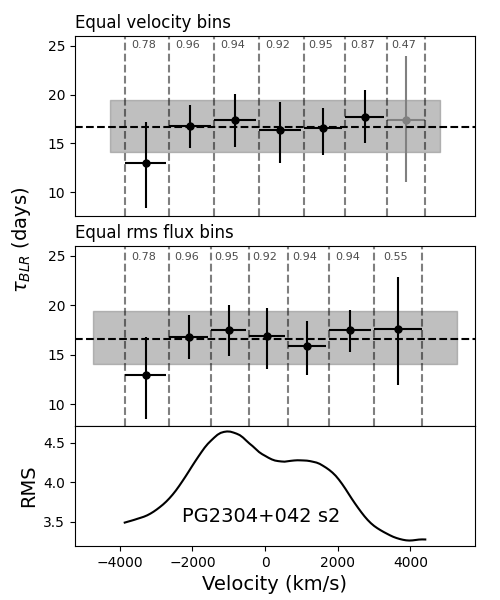}{0.3\textwidth}{}
        \fig{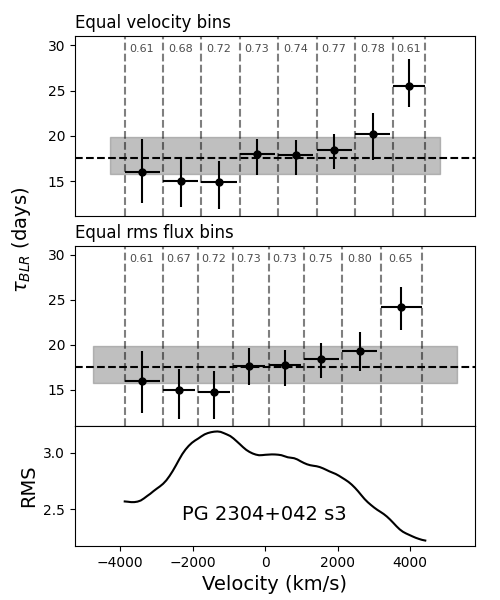}{0.3\textwidth}{}
        \fig{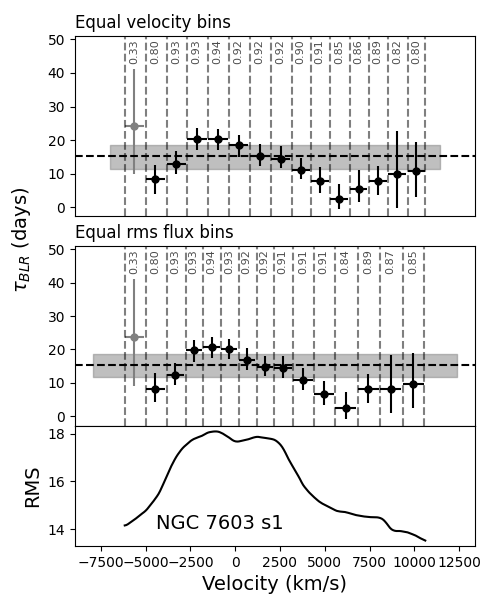}{0.3\textwidth}{}}
\caption{Velocity-resolved lags. The meanings of the panels are the same as in Figure~\ref{figure:v_resolved1}}
\label{figure:v_resolved4}
\end{figure*}
\end{center}

We used two different binning methods to calculate time lags as a function of velocity: equal velocity bins and equal rms flux density bins. The advantage of equal rms flux density bins is that they allot similar noise to each bin. We present both binning methods, as they should roughly agree when robust results are obtained. We divided each object's H$\beta$ profile into as many bins as fit across the line, while constraining the bins to remain larger than our effective spectral resolution of 925 km s$^{-1}$. This ranges from 6 to 15 bins. We remeasured the binned flux densities using the integration method for each epoch and then cross-correlated each bin with the continuum, building up a CCCD for each bin. We used non-narrow-line subtracted spectra for the velocity-resolved analysis. The velocity-resolved time lags are plotted in Figures~\ref{figure:v_resolved1}--\ref{figure:v_resolved4}. Table~\ref{tab:vel_asymm} lists the number of bins, kinematics of the BLR, and line asymmetry, with further discussion on individual objects in Section~\ref{sec:dis}.

\begin{deluxetable*}{lcccccccc}
\tablecaption{Line Widths and Time Lags \label{tab:widthlags}}
\tablewidth{0pt}
\tablehead{
\colhead{} & \colhead{} & \multicolumn2c{rms} & \multicolumn2c{mean} & \multicolumn3c{Time Lags} 
\\ \cline{3-4} \cline{5-6} \cline{7-9}
\colhead{Object} & \colhead{Season} & \colhead{\textbf{$\sigma_{\textrm{line}}$}} & \colhead{FWHM} & \colhead{$\sigma_{\textrm{line}}$} & \colhead{FWHM} & \colhead{\textbf{$\tau_{\textrm{cent}}$}} & \colhead{$\tau_{\textrm{peak}}$} & \colhead{Max lag}\\
\colhead{} & \colhead{} & \colhead{km s$^{-1}$} & \colhead{km s$^{-1}$} & \colhead{km s$^{-1}$} & \colhead{km s$^{-1}$} & \colhead{days} & \colhead{days} & \colhead{days}\\
\colhead{(1)} & \colhead{(2)} & \colhead{(3)} & \colhead{(4)} & \colhead{(5)} & \colhead{(6)} & \colhead{(7)} & \colhead{(8)} & \colhead{(9)}
}

\startdata
1ES 0206+522 & 1 & 1898 $\pm$ 100 & 6291 $\pm$ 662 & 1471 $\pm$ 100 & 3178 $\pm$ 99 & 28.4 $\pm ^{{9.1}} _{{7.3}}$ & 22.5 $\pm ^{{4.5}} _{{17.3}}$ & 80 \\ 
& 2 & 928 $\pm$ 77 & 2905 $\pm$ 88 & 1473 $\pm$ 104 & 3359 $\pm$ 93 & 54.2 $\pm ^{{4.8}} _{{7.7}}$ & 53.6  $\pm ^{{9.2}} _{{10.0}}$ & 105 \\ 
& 5 & 1187 $\pm$ 92 & 2815 $\pm$ 91 & 1269 $\pm$ 94 & 2631 $\pm$ 87 & 35.6 $\pm ^{{5.2}} _{{6.1}}$ & 40.3 $\pm ^{{8.2}} _{{4.4}}$ & 80 \\ 
Mrk 1040 & 1 & 1194 $\pm$ 94 & 3107 $\pm$ 102 & 1312 $\pm$ 98 & 3575 $\pm$ 95 & 5.6 $\pm ^{{1.4}} _{{1.5}}$ & 7.6 $\pm ^{{4.8}} _{{1.7}}$ & 25 \\ 
& 3 & 1074 $\pm$ 98 & 2831 $\pm$ 94 & 1324 $\pm$ 95 & 3396 $\pm$ 99 & 21.0 $\pm ^{{7.1}} _{{5.3}}$ & 14.3 $\pm ^{{7.6}} _{{7.3}}$ & 60 \\ 
Mrk 618 & 1 & 1650 $\pm$ 121 & 3219 $\pm$ 82 & 1574 $\pm$ 123 & 2758 $\pm$ 94 & 9.2 $\pm ^{{1.6}} _{{2.3}}$ & 10.9 $\pm ^{{2.7}} _{{1.6}}$ & 30 \\ 
& 2 & 1573 $\pm$ 134 & 1903 $\pm$ 270 & 1525 $\pm$ 118 & 2758 $\pm$ 93 & 15.2 $\pm ^{{2.4}} _{{2.3}}$ & 14.9 $\pm ^{{2.3}} _{{5.4}}$ & 60 \\ 
& 3 & 1492 $\pm$ 110 & 3494 $\pm$ 95 & 1490 $\pm$ 119 & 2758 $\pm$ 87 & 30.9 $\pm ^{{10.6}} _{{7.2}}$ & 28.1 $\pm ^{{6.9}} _{{12.5}}$ & 60 \\ 
& 4 & 1279 $\pm$ 104 & 2387 $\pm$ 96 & 1465 $\pm$ 118 & 2763 $\pm$ 98 & 10.4 $\pm ^{{2.8}} _{{2.9}}$ & 11.0 $\pm ^{{4.8}} _{{5.2}}$ & 50 \\
MCG -02-14-009 & 3 & 1784 $\pm$ 100 & 4584 $\pm$ 126 & 1704 $\pm$ 98 & 4494 $\pm$ 92 & 12.1 $\pm ^{{4.2}} _{{4.7}}$ & 13.5$\pm ^{{8.8}} _{{4.9}}$ & 40 \\
& 4 & 1821 $\pm$ 106 & 3406 $\pm$ 92 & 1587 $\pm$ 95 & 4404 $\pm$ 90 & 11.5 $\pm ^{{7.4}} _{{9.9}}$ & 10.2 $\pm ^{{10.1}} _{{15.1}}$ & 40\\
& 5 & 1713 $\pm$ 106 & 3424 $\pm$ 99 & 1695 $\pm$ 98 & 4245 $\pm$ 91 & 17.4 $\pm ^{{5.8}} _{{11.9}}$ & 10.6 $\pm ^{{2.9}} _{{4.9}}$ & 75\\
IRAS 05589+2828 & 1 & 2455 $\pm$ 112 & 6924 $\pm$ 96 & 2669 $\pm$ 125 & 5583 $\pm$ 90 & 29.0 $\pm ^{{10.6}} _{{6.8}}$ & 18.9 $\pm ^{{3.9}} _{{24.5}}$ & 80 \\
& 2 & 2784 $\pm$ 128 & 4865 $\pm$ 1072 & 2633 $\pm$ 128 & 4955 $\pm$ 90 & 19.0 $\pm ^{{5.5}} _{{4.7}}$ & 21.8 $\pm ^{{5.1}} _{{2.9}}$ & 60 \\
& 3 & 2776 $\pm$ 178 & 4142 $\pm$ 92 & 2616 $\pm$ 132 & 5044 $\pm$ 91 & 21.0 $\pm ^{{2.0}} _{{2.5}}$ & 21.2 $\pm ^{{1.9}} _{{2.9}}$ & 50 \\
& 4 & 2210 $\pm$ 126 & 5314 $\pm$ 91 & 2627 $\pm$ 135 & 5314 $\pm$ 96 & 44.7 $\pm ^{{3.4}} _{{3.2}}$ & 48.0 $\pm ^{{5.0}} _{{1.9}}$ & 100 \\
& 5 & 2360 $\pm$ 117 & 5330 $\pm$ 508 & 2576 $\pm$ 131 & 5330 $\pm$ 95 & 38.4 $\pm ^{{5.0}} _{{4.7}}$ & 33.8 $\pm ^{{4.1}} _{{12.6}}$  & 80 \\
Mrk 715 & 4 & 3967 $\pm$ 174 & 10691 $\pm$ 2307 & 4306 $\pm$ 163 & 11878 $\pm$ 140 & 32.5 $\pm ^{{8.1}} _{{9.5}}$ & 34.6 $\pm ^{{10.6}} _{{9.8}}$ & 70 \\
& 5 & 3659 $\pm$ 161 & 6013 $\pm$ 144 & 4431 $\pm$ 163 & 12894 $\pm$ 120 & 17.8 $\pm ^{{7.6}} _{{9.8}}$ & 15.7 $\pm ^{{9.1}} _{{20.4}}$ & 70 \\
SBS 1136+594 & 1 & 1207 $\pm$ 76 & 3587 $\pm$ 85 & 1504 $\pm$ 96 & 3765 $\pm$ 86 & 9.8 $\pm ^{{2.1}} _{{2.8}}$ & 8.7 $\pm ^{{2.1}} _{{4.1}}$ & 40 \\
& 2 & 1364 $\pm$ 94 & 2978 $\pm$ 88 & 1556 $\pm$ 103 & 4056 $\pm$ 99 & 21.3 $\pm ^{{5.7}} _{{7.1}}$ & 19.0 $\pm ^{{4.0}} _{{11.1}}$ & 60 \\
VIII Zw 233 & 2 & 1731 $\pm$ 82 & 4828 $\pm$ 303 & 1861 $\pm$ 104 & 5161 $\pm$ 141 & 21.9 $\pm ^{{5.4}} _{{6.7}}$ & 20.7 $\pm ^{{2.8}} _{{11.1}}$ & 60 \\
& 3 & 2020 $\pm$ 102 & 1735 $\pm$ 1003 & 1904 $\pm$ 101 & 5411 $\pm$ 165 & 19.4 $\pm ^{{4.1}} _{{6.6}}$ & 20.1 $\pm ^{{8.9}} _{{7.6}}$ & 60 \\
Mrk 813 & 4 & 1748 $\pm$ 102 & 4793 $\pm$ 82 & 1970 $\pm$ 104 & 5884 $\pm$ 88 & 33.8 $\pm ^{{14.2}} _{{11.5}}$ & 39.0 $\pm ^{{26.5}} _{{8.7}}$ & 80 \\ 
SDSS J145307.92+255433.0 & 1 & 4095 $\pm$ 171 & 8573 $\pm$ 1829 & 4235 $\pm$ 168 & 12548 $\pm$ 95 & 23.2 $\pm ^{{5.5}} _{{5.9}}$ & 24.0 $\pm ^{{10.7}} _{{7.5}}$ & 60 \\ 
SDSS J152139.66+033729.2 & 3 & 4099 $\pm$ 138 & 11252 $\pm$ 444 & 4250 $\pm$ 160 & 7742 $\pm$ 83 & 20.0 $\pm ^{{2.6}} _{{4.8}}$ & 16.9 $\pm ^{{3.9}} _{{5.7}}$ & 60 \\ 
2MASX J21090996-0940147 & 1 & 935 $\pm$ 90 & 2700 $\pm$ 88 & 1376 $\pm$ 94 & 3538 $\pm$ 88 & 6.5 $\pm ^{{2.2}} _{{2.6}}$ & 4.9 $\pm ^{{2.8}} _{{5.4}}$ & 25\\ 
& 3 & 1017 $\pm$ 83 & 2700 $\pm$ 87 & 1366 $\pm$ 96 & 3168 $\pm$ 94 & 16.3 $\pm ^{{3.1}} _{{2.3}}$ & 13.6 $\pm ^{{1.9}} _{{2.6}}$ & 50 \\ 
& 4 & 925 $\pm$ 89 & 2513 $\pm$ 91 & 1233 $\pm$ 99 & 2796 $\pm$ 96 & 29.6 $\pm ^{{6.5}} _{{5.2}}$ & 24.2 $\pm ^{{3.2}} _{{15.8}}$ & 80 \\
PG 2304+042 & 1 & 1484 $\pm$ 79 & 4929 $\pm$ 1061 & 1726 $\pm$ 100 & 5733 $\pm$ 84 & 17.6 $\pm ^{{2.3}} _{{1.8}}$ & 17.1 $\pm ^{{2.9}} _{{5.2}}$ & 50 \\ 
& 2 & 1496 $\pm$ 86 & 4750 $\pm$ 85 & 1728 $\pm$ 91 & 5733 $\pm$ 89 & 16.6 $\pm ^{{2.8}} _{{2.5}}$ & 16.9 $\pm ^{{3.9}} _{{2.6}}$ & 60 \\
& 3 & 1580 $\pm$ 87 & 5295 $\pm$ 89 & 1723 $\pm$ 103 & 5563 $\pm$ 94 & 17.6 $\pm ^{{2.3}} _{{1.8}}$ & 16.8 $\pm ^{{1.6}} _{{4.3}}$ & 60 \\
NGC 7603 & 1 & 3508 $\pm$ 157 & 7839 $\pm$ 95 & 4130 $\pm$ 177 & 9644 $\pm$ 99 & 15.4 $\pm ^{{3.2}} _{{3.8}}$ & 11.5 $\pm ^{{4.3}} _{{5.7}}$ & 60 
\enddata

\tablecomments{Time lags and line widths are in the rest frame. Columns 3 \& 4 are the line dispersion and FWHM measured from the rms profile respectively. Columns 5 \& 6 are the line dispersion and FWHM measured from the mean profile respectively. Column 7 is the time lag measured from the centroid of the CCF. Column 8 is the time lag measured from the peak of the CCF. Column 9 is the maximum time lag searched. We bold the header of columns 3 and 7 as they list our preferred line width and time lag measurements.}
\end{deluxetable*}

\begin{deluxetable*}{!{\extracolsep{1pt}}lcccccc}
\tablecaption{Virial Products and Black Hole Masses \label{tab:vp_mass}}
\tablewidth{0pt}
\tablehead{
\colhead{} & \colhead{} & \multicolumn{1}{c}{Virial Product} & \multicolumn{2}{c}{Black Hole Mass}
\\ \cline{3-5}  
\colhead{Object} & \colhead{Season} & \colhead{R$_{\textrm{H}\beta}$V$^2$$_{\textrm{FWHM}}$/G} & \colhead{1.12 $\times$ R$_{\textrm{H}\beta}$V$^2$$_{\textrm{FWHM}}$/G} & \colhead{\textbf{4.47 $\times$ R$_{\textrm{H}\beta}$$\sigma^2_{\textrm{line}}$/G}} & \colhead{Notes}
\\ 
\colhead{} & \colhead{} & \colhead{$\times$10$^7$M$_\odot$} & \colhead{$\times$10$^7$M$_\odot$} & \colhead{$\times$10$^7$M$_\odot$}\\
\colhead{(1)} & \colhead{(2)} & \colhead{(3)} & \colhead{(4)} & \colhead{(5)} & \colhead{(6)}
}

\startdata
1ES 0206+522 & 1 & 5.6 $\pm ^{{2.0}} _{{1.6}}$ & 24.5 $\pm ^{{9.4}} _{{8.2}}$ & 8.9 $\pm ^{{3.0}} _{{2.5}}$ & \\ 
& 2 & 11.9 $\pm ^{{1.4}} _{{2.1}}$ & 10.0 $\pm ^{{1.1}} _{{1.6}}$ & 4.1 $\pm ^{{0.77}} _{{0.90}}$ & \checkmark \\ 
& 5 & 4.8 $\pm ^{{0.87}} _{{0.99}}$ & 6.2 $\pm ^{{0.99}} _{{1.1}}$ & 4.4 $\pm ^{{0.93}} _{{1.0}}$ & \\ 
Mrk 1040 & 1 & 1.4 $\pm ^{{0.40}} _{{0.42}}$ & 1.2 $\pm ^{{0.30}} _{{0.32}}$ & 0.70 $\pm ^{{0.20}} _{{0.21}}$ & \checkmark \\
& 3 & 4.7 $\pm ^{{1.8}} _{{1.4}}$ & 3.7 $\pm ^{{1.3}} _{{1.0}}$ & 2.1 $\pm ^{{0.81}} _{{0.65}}$ & \\ 
Mrk 618 & 1 & 1.4 $\pm ^{{0.29}} _{{0.40}}$ & 2.1 $\pm ^{{0.38}} _{{0.53}}$ & 2.2 $\pm ^{{0.50}} _{{0.63}}$ & \\
& 2 & 2.3 $\pm ^{{0.43}} _{{0.41}}$ & 1.2 $\pm ^{{0.39}} _{{0.39}}$ & 3.3 $\pm ^{{0.76}} _{{0.74}}$ & \\ 
& 3 & 4.6 $\pm ^{{1.8}} _{{1.2}}$ & 8.2 $\pm ^{{2.9}} _{{2.0}}$ & 6.0 $\pm ^{{2.2}} _{{1.7}}$ & \\ 
& 4 & 1.6 $\pm ^{{0.49}} _{{0.50}}$ & 1.3 $\pm ^{{0.37}} _{{0.38}}$ & 1.5 $\pm ^{{0.47}} _{{0.48}}$ & \checkmark \\ 
MCG -02-14-009 & 3 & 4.8 $\pm ^{{1.9}} _{{2.1}}$ & 5.6 $\pm ^{{2.0}} _{{2.2}}$ & 3.4 $\pm ^{{1.2}} _{{1.4}}$ & \checkmark \\
& 4 & 4.3 $\pm ^{{3.1}} _{{4.2}}$ & 2.9 $\pm ^{{1.9}} _{{2.5}}$ & 3.3 $\pm ^{{2.2}} _{{2.9}}$ & \\ 
& 5 & 6.1 $\pm ^{{2.3}} _{{4.7}}$ & 4.5 $\pm ^{{1.5}} _{{3.1}}$ & 4.5 $\pm ^{{1.6}} _{{3.1}}$ & \\ 
IRAS 05589+2828 & 1 & 17.6 $\pm ^{{7.2}} _{{4.7}}$ & 30.4 $\pm ^{{11.1}} _{{7.2}}$ & 15.2 $\pm ^{{5.7}} _{{3.8}}$ &  \\ 
& 2 & 9.1 $\pm ^{{3.0}} _{{2.6}}$ & 9.8 $\pm ^{{5.2}} _{{5.0}}$ & 12.9 $\pm ^{{3.9}} _{{3.4}}$ &  \\ 
& 3 & 10.4 $\pm ^{{1.2}} _{{1.5}}$ & 7.9 $\pm ^{{0.84}} _{{1.0}}$ & 14.1 $\pm ^{{2.3}} _{{2.5}}$ & \checkmark \\ 
& 4 & 24.6 $\pm ^{{2.3}} _{{2.2}}$ & 27.6 $\pm ^{{2.3}} _{{2.2}}$ & 19.0 $\pm ^{{2.6}} _{{2.6}}$ & \\ 
& 5 & 21.3 $\pm ^{{3.2}} _{{3.1}}$ & 23.8 $\pm ^{{5.5}} _{{5.4}}$ &  18.7 $\pm ^{{3.0}} _{{3.0}}$ & \\ 
Mrk 715 & 4 & 89.4 $\pm ^{{25.0}} _{{29.3}}$ & 81.1 $\pm ^{{40.4}} _{{42.2}}$ & 44.6 $\pm ^{{11.7}} _{{13.6}}$ & \\ 
& 5 & 57.7 $\pm ^{{27.5}} _{{35.6}}$ & 14.0 $\pm ^{{6.0}} _{{7.8}}$ & 20.8 $\pm ^{{9.0}} _{{11.6}}$ & \checkmark \\ 
SBS 1136+594 & 1 & 2.7 $\pm ^{{0.67}} _{{0.88}}$ & 2.8 $\pm ^{{0.61}} _{{0.80}}$ & 1.2 $\pm ^{{0.32}} _{{0.40}}$ & \checkmark \\
& 2 & 6.9 $\pm ^{{2.1}} _{{2.6}}$ & 4.2 $\pm ^{{1.1}} _{{1.4}}$ & 3.5 $\pm ^{{1.0}} _{{1.2}}$ & \\ 
VIII Zw 233 & 2 & 11.4 $\pm ^{{3.2}} _{{4.0}}$ & 11.1 $\pm ^{{3.1}} _{{3.7}}$ & 5.7 $\pm ^{{1.5}} _{{1.8}}$ & \checkmark \\ 
& 3 & 11.1 $\pm ^{{2.7}} _{{4.3}}$ & 1.3 $\pm ^{{1.5}} _{{1.5}}$ & 6.7 $\pm ^{{1.6}} _{{2.4}}$ & \\ 
Mrk 813 & 4 & 22.9 $\pm ^{{10.8}} _{{8.7}}$ & 17.0 $\pm ^{{7.1}} _{{5.8}}$ & 9.0 $\pm ^{{3.9}} _{{3.2}}$ & \checkmark \\ 
SDSS J145307.92+255433.0.0 & 1 & 71.4 $\pm ^{{19.0}} _{{20.4}}$ & 37.3 $\pm ^{{18.2}} _{{18.5}}$ & 34.0 $\pm ^{{8.5}} _{{9.1}}$ & \checkmark \\ 
SDSS J152139.66+033729.2 & 3 & 23.4 $\pm ^{{3.4}} _{{6.3}}$ & 55.4 $\pm ^{{8.4}} _{{14.0}}$ & 29.3 $\pm ^{{4.3}} _{{7.3}}$ & \checkmark \\ 
2MASX J21090996-0940147 & 1 & 1.6 $\pm ^{{0.61}} _{{0.72}}$ & 1.0 $\pm ^{{0.36}} _{{0.42}}$ & 0.50 $\pm ^{{0.19}} _{{0.22}}$ & \checkmark \\ 
& 3 & 3.2 $\pm ^{{0.71}} _{{0.55}}$ & 2.6 $\pm ^{{0.52}} _{{0.41}}$ & 1.5 $\pm ^{{0.37}} _{{0.32}}$ & \\ 
& 4 & 4.5 $\pm ^{{1.2}} _{{0.96}}$ & 4.1 $\pm ^{{0.94}} _{{0.78}}$ & 2.2 $\pm ^{{0.64}} _{{0.57}}$ & \\ 
PG 2304+042 & 1 & 10.9 $\pm ^{{2.1}} _{{3.8}}$ & 9.1 $\pm ^{{4.2}} _{{4.8}}$ & 3.3 $\pm ^{{0.65}} _{{1.1}}$ &  \\ 
& 2 & 10.7 $\pm ^{{2.0}} _{{1.9}}$ & 8.2 $\pm ^{{1.4}} _{{1.3}}$ & 3.3  $\pm ^{{0.66}} _{{0.62}}$ & \\ 
& 3 & 10.6 $\pm ^{{1.6}} _{{1.3}}$ & 10.8 $\pm ^{{1.5}} _{{1.2}}$ & 3.8 $\pm ^{{0.65}} _{{0.58}}$ & \checkmark \\ 
NGC 7603 & 1 & 27.9 $\pm ^{{6.6}} _{{7.8}}$ & 20.7 $\pm ^{{4.4}} _{{5.2}}$ & 16.5 $\pm ^{{3.8}} _{{4.4}}$ & \checkmark  
\enddata
\tablecomments{The virial product uses line-width measurements from the mean profiles. The mass measurements use line-widths measured from the rms profiles. The check marks in column 6 denote our preferred mass measurement for an individual object. We bold the header of column 5 as we prefer the mass measurement using $\sigma^2_{\textrm{line}}$ measured from the rms profile.}
\end{deluxetable*}

\begin{deluxetable*}{lccccc}
\tablecaption{Velocity-Resolved Results and H$\beta$ Asymmetry \label{tab:vel_asymm}}
\tablewidth{0pt}
\tablehead{
\colhead{Object} & \colhead{Season} & \colhead{Velocity Bins} & \colhead{BLR Kinematics} & \colhead{H$\beta$ Asymmetry} & \colhead{H$\beta$ range}\\
\colhead{} & \colhead{} & \colhead{} & \colhead{} & \colhead{} & \colhead{\r{A}}\\
\colhead{(1)} & \colhead{(2)} & \colhead{(3)} & \colhead{(4)} & \colhead{(5)} & \colhead{(6)}
}

\startdata
1ES 0206+522 & 1 & 8 & Disk-like & 0.014 & \\
& 2 & 6 & Flat & 0.026 & 4800-4915\\
& 5 & 6 & Disk-like & 0.066 & \\
Mrk 1040 & 1 & 6 & Disk-like & -0.088 & \\
& 3 & 6 & Outflow & -0.146 & \\
Mrk 618 & 1 & 7 & Indeterminate & -0.256 & \\
& 2 & 7 & Disk-like & -0.192 & \\
& 3 & 7 & Indeterminate & -0.144 & \\
& 4 & 6 & Outflow & -0.160 & \\
MCG -02-14-009 & 3 & 8 & Indeterminate & -0.049 & 4791-4930\\
& 4 & 8 & Indeterminate & -0.080 & \\
& 5 & 8 & Indeterminate & -0.084 & \\
IRAS 05589+2828 & 1 & 11 & Disk-like & -0.095 & \\
& 2 & 11 & Inflow & -0.107 & \\
& 3 & 11 & Indeterminate & -0.114 & \\
& 4 & 11 & Indeterminate & -0.100 & \\
& 5 & 11 & Indeterminate & -0.075 & \\
Mrk 715 & 4 & 15 & Outflow & -0.121 & \\
& 5 & 15 & Disk-like & -0.233 & \\
SBS 1136+594 & 1 & 6 & Indeterminate & -0.114 & \\
& 2 & 7 & Outflow & -0.075 & \\
VIII Zw 233 & 2 & 8 & Disk-like & 0.121 & \\
& 3 & 8 & Disk-like & 0.121 & \\
Mrk 813 & 4 & 8 & Inflow & -0.028 & \\
SDSS J145307.92+255433.0.0 & 1 & 13 & Disk-like & -0.081 & 4770-4990 \\
SDSS J152139.66+033729.2 & 3 & 12 & Disk-like & -0.236 & \\
2MASX J21090996-0940147 & 1 & 6 & Indeterminate & 0.013 & \\
& 3 & 6 & Disk-like & -0.028 & \\
& 4 & 6 & Indeterminate & 0.032 & \\
PG 2304+042 & 1 & 7 & Disk-like & -0.054 & \\
& 2 & 7 & Flat & -0.038 & \\
& 3 & 8 & Outflow & -0.040 & \\
NGC 7603 & 1 & 15 & Disk-like & -0.214 & 4760-5034
\enddata

\tablecomments{Column 1 lists the object name. Column two lists the season. Column 3 lists the number of velocity bins. Column 4 lists the BLR dynamics as indicated by the velocity resolved time lags. Column 5 lists the asymmetry where negative numbers indicate red asymmetry and positive numbers indicate blue asymmetry. Column 5 lists the H$\beta$ integration range when it is different than the range used for constructing the light curves in Table~\ref{tab:OIII}. Wavelengths are in the rest frame.}
\end{deluxetable*}

\section{Discussion} \label{sec:dis}
Below we discuss results for individual objects.
\subsection{1ES 0206+522} \label{subsec:1es}
See the light curve, CCFs, CCCDs, mean, and rms spectra for this object in Figure~\ref{figure:1es_lc}. We selected this object for MAHA based on its asymmetric H$\beta$ profile observed in 2012 (private communication, WIRO). Although this object showed a significantly less asymmetric profile at the beginning of MAHA, we included it in the campaign as it is bright, observable nearly year-round at WIRO, and historically showed asymmetric H$\beta$. We observed this object for five seasons beginning in 2018.
In the most recent season (season 5), 1ES 0206+522 started to show stronger blue asymmetry.  We measured a time lag in three of those seasons, albeit changing lags across all three seasons, with season 1 showing the shortest and season 2 the longest lag. Season 1 has the largest mass measurement (\textrm{$8.9 \times 10^{7} M_\odot$}) and season 2 has the smallest (\textrm{$4.1 \times 10^{7} M_\odot$}, which is our preferred mass). 
We detrended the season 1 continuum light curve, noting that the continuum light curve shows a long-term decreasing trend while the H$\beta$ light curve remains mostly flat until the last few epochs of the season. The time lag, CCF, and CCCD for season 1 are measured from the detrended light curve.
The velocity-resolved lags indicate changing kinematics in different seasons. Season 1 and 5 suggest disk-like structure. Season 2 has a nearly constant lag across all velocities (see Figure~\ref{figure:v_resolved1}).
We excluded seasons 3 \& 4 as we did not observe a clear reverberation.

\subsection{Mrk 1040} \label{subsec:mrk1040}
See the light curve, CCFs, CCCDs, mean, and rms spectra for this object in Figure~\ref{figure:mrk1040_lc}. Mrk 1040 is a bright Seyfert galaxy without particularly strong H$\beta$ asymmetry in our first season, and was initially selected to fill a gap in the observing schedule near the Galactic plane during long winter nights.  We dropped Mrk 1040 after one season to make room for targets with more asymmetric lines, but later resumed to obtain a second epoch of reverberation mapping to look for any changes in the H$\beta$ profile and any corresponding changes in the BLR kinematics.
Seasons 1 \& 3 provide time lags, while season 2 is inconclusive. The CCF for season 3 is much broader than for season 1, showing a large ``shoulder" for lags longer than the measured lag. This is likely because the continuum light curve shows two inflection points, and the H$\beta$ light curve could be matched up with either inflection point. The mass measured in season 3 is larger than that measured in season 1 (\textrm{$2.1 \times 10^{7} M_\odot$} and \textrm{$7.0 \times 10^{6} M_\odot$} respectively). The season 1 mass is our preferred mass.
Aside from the first bin, the velocity-resolved lags for season 1 show a general disk-like pattern. 
The velocity-resolved lags for season 3 indicate outflow (see Figure~\ref{figure:v_resolved1}). The asymmetry is red in both seasons, though it is stronger in season 3.

\subsection{Mrk 618} \label{subsec:mrk618}
See the light curve, CCFs, CCCDs, mean, and rms spectra for this object in Figure!\ref{figure:mrk618_lc}. Mrk 618 is a bright Seyfert galaxy with asymmetric H$\beta$ and is notably a potential GRAVITY target given its southern declination \citep{SARM}.
Mrk 618 was observed as part of a 2012 \citet{DeRosa2018} RM campaign, but they did not obtain a robust measurement of an H$\beta$ time lag. We have been observing this object since 2019 and have four seasons of data. We measure a time lag from all four seasons. 
Different time lags are measured for the four seasons, ranging from 9.2 to 30.9 days. Season 2 displays the strongest variability and clearest reverberation.
The CCF for season 3 is comparatively broad and shows two peaks, making the season 3 lag measurement the most uncertain. 
The black hole masses from seasons 1 \& 2 and 1 \& 4 agree to within the measurement uncertainty. The season 4 mass is our preferred mass.
The velocity-resolved lags for season 2 looks generally disk-like, while the seasons 4 kinematics indicate outflow. The velocity-resolved lags for seasons 1 \& 3 are indeterminate (see Figure~\ref{figure:v_resolved1}). All four seasons show red asymmetry with season 1 showing the strongest asymmetry.

\subsection{MCG -02-14-009} \label{subsec:mcg}
See the light curve, CCFs, CCCDs, mean, and rms spectra for this object in Figure~\ref{figure:mcg_lc}. This target is another bright southern Seyfert galaxy that may be a good target for GRAVITY \citep{SARM}.
We have been observing MCG -02-14-009 since 2019 and have five seasons of data. Seasons 1 \& 2 are non-results due to a low number of epochs. We measured similar time lags for seasons 3 \& 4 (12.1 \& 11.5 days respectively) and a longer lag for season 5 (17.4 days). The uncertainties on these lags are large and all three agree to within the measurement uncertainties.
The season 5 CCF is double peaked with a broad CCCD. As a result, the season 5 lag is the most uncertain. The three mass measurements agree to within the margin of error. The season 3 mass is our preferred mass.
The velocity-resolved time lags are indeterminate for all three seasons (see Figure~\ref{figure:v_resolved2}). All three seasons show red asymmetry, with season 5 having the strongest red asymmetry.

\subsection{IRAS 05589+2828} \label{subsec:iras}
See the light curve, CCFs, CCCDs, mean, and rms spectra for this object in Figure~\ref{figure:iras_lc}. We initially targeted this object as its H$\beta$ was described as having ``a red wing'' \citep{Winter2010}.
This object is part of a close dual AGN with companion 2MASX J06021107+2828382 at a separation of 8 kpc \citep[$\sim12$'';][]{Koss2012,Benitez2022}. We have five seasons of data on IRAS 05589+2828, beginning in 2018. This object shows clear variation in each season and also shows a significant flux density and flux drop in season 3 with a sharp flux density and flux rise in season 4. We are able to measure time lags for each season. The lags are similar in seasons 1-3, about 20 days, and increase in seasons 4 \& 5 to about 40 and 33 days, respectively. The mass measurements from seasons 1-5 agree within the uncertainties. The season 3 mass is our preferred mass.
The velocity-resolved results for this object show a disk-like structure for season 1, and evidence of inflow for season 2. The velocity-resolved lags for Seasons 3-5 are indeterminate (see Figure~\ref{figure:v_resolved2}). All seasons show red asymmetry with season 3 showing the strongest red asymmetry.

\subsection{Mrk 715} \label{subsec:mrk715}
See the light curve, CCFs, CCCDs, mean, and rms spectra for this object in Figure~\ref{figure:sdss1004_lc}. We described this object, also known as SDSS J100447.61+144645.6, as having ``a double-peaked H$\beta$ line profile and a long tail to the red" \citep{MAHA1}.
We have six seasons of data on Mrk 715, but were only able to measure time lags in seasons 4 \& 5. Seasons 1-3 \& 6 did not have favorable variability and our temporal coverage was not extensive. The variable component of H$\beta$ in the rms spectra of this object is weak and appears dominated by a variable continuum. We followed MAHA III in fitting an AGN power-law continuum to the spectra using a non-linear least-squares minimization fit and subtracting the AGN power-law continuum from each epoch of spectra before constructing the rms spectra for both season 4 and season 5. The time lag in season 4 is almost twice that of the time lag in season 5. Correspondingly, the season 4 mass is more than twice that of the season 5 mass given the similar line widths. The season 5 mass is our preferred mass as it has smaller measurement uncertainty. 
The velocity-resolved results for season 4 shows outflow, while season 5 shows a mostly disk-like structure, though the rms profiles are noisy and we caution against giving these velocity-resolved measurements too much weight (see Figure~\ref{figure:v_resolved2} and Figure~\ref{figure:v_resolved3}). Both seasons show red asymmetry with season 5 having the stronger asymmetry.

\subsection{SBS 1136+594} \label{subsec:sbs}
See the light curve, CCFs, CCCDs, mean, and rms spectra for this object in Figure~\ref{figure:sbs_lc}. This object was selected based on the asymmetric profile published by \citet{Zamfir2010} that displays bumps on both the blue and red wings, with excess emission on the red side.
We have two seasons of data, both showing a strong inflection point in the continuum with a clear, delayed response in H$\beta$. In season 2 we measure a lag more than twice that in season 1, though they agree to within the uncertainties. The CCCD for season 2 is far broader than the CCCD for season 1, leading to relatively larger uncertainties in the season 2 lag and mass measurements. Season 1 is our preferred mass. The velocity-resolved results are indeterminate in season 1 and indicate outflow in season 2 (see Figure~\ref{figure:v_resolved3}). Both seasons show $H\beta$ having red asymmetry with season 1 having the stronger asymmetry.

\subsection{VIII Zw 233} \label{subsec:viiizw}
See the light curve, CCFs, CCCDs, mean, and rms spectra for this object in Figure~\ref{figure:viiizw233_lc}. We described this object as having ``a redshifted peak" \citep{MAHA1}. We have been observing this object for five seasons, since 2016. Only seasons 2 \& 3 produced a measurable time lag. Season 2 shows an apparent flare in the continuum light curve and a strong response a short time later from H$\beta$. The variable component of H$\beta$ in the rms spectra of this object is weak and appears dominated by a variable continuum. We follow MAHA III in fitting an AGN power-law continuum to the spectra using a non-linear leas-squares minimization fit and subtracting the AGN power-law continuum from each epoch of spectra before constructing the rms spectra for both season 2 and season 3. After calculating this continuum subtracted rms spectrum, it is clear that season 2 has much stronger variability in H$\beta$ than season 3. The season 2 mass is our preferred mass. The velocity-resolved lags show a disk-like structure in both seasons (see Figure~\ref{figure:v_resolved3}). Both seasons show blue asymmetry.

\subsection{Mrk 813} \label{subsec:mrk813}
See the light curve, CCFs, CCCDs, mean, and rms spectra for this object in Figure~\ref{subsec:mrk813}. This object is a bright Seyfert galaxy that shows a bump and excess emission on the red side of the H$\beta$ profile in SDSS spectra, although the line appears more symmetric in our most recent spectra.
We have four seasons of data on this object, but not every year as the priority was dropped after poor initial results. Reprioritizing Mrk 813 in 2022 resulted in finally measuring a time lag, although the situation is not totally straightforward. The CCF for season 4 is a double peaked which leads to a broad CCCD. The two peaks are separated by approximately 40 days. The two peaks are not strongly different than a broad, singly peaked CCF, and so we trust our CCCD which indicates a lag in the center of the two peaks. This season shows red asymmetry.
The velocity-resolved results for this object show inflow (see Figure~\ref{figure:v_resolved3}). 

\subsection{SDSS J145307.92+255433.0.0} \label{subsec:sdssj1453}
See the light curve, CCFs, CCCDs, mean, and rms spectra for this object in Figure~\ref{figure:sdssj1453_lc}. This object has a strange H$\beta$ profile in a SDSS spectrum and also appeared double-peaked and asymmetric in our WIRO data.
We only have one season of observations for this object and ZTF photometry helps significantly in covering gaps in the continuum light curve. The variation in our single season is strong and echoed by the H$\beta$. This season shows red asymmetry.
The velocity-resolved results for this object show a likely disk-like structure (see Figure~\ref{figure:v_resolved3}).

\subsection{SDSS J152139.66+033729.2} \label{subsec:sdss1521}
See the light curve, CCFs, CCCDs, mean, and rms spectra for this object in Figure~\ref{figure:sdss1521_lc}. We previously described this object as having ``an H$\beta$ line with a red asymmetry" \citep{MAHA1}.
We have four seasons of this object, extending back to 2017, but only observed an inflection point in both the continuum and H$\beta$ in season 3, allowing us to measure a time lag. The lag in season 3 is clear, and the velocity-resolved results indicate a disk-like structure (see Figure~\ref{figure:v_resolved3}). This season shows red asymmetry.

\subsection{2MASX J21090996-0940147} \label{subsec:2masx}
See the light curve, CCFs, CCCDs, mean, and rms spectra for this object in Figure~\ref{figure:2masx_lc}. This target is another bright southern object that may be a good target for GRAVITY \citep{SARM}. 
We have four seasons of data on this object, beginning in 2019, but only measure a lag in three seasons (1 \& 3-4). Our measured lag is shortest in season 1 and longest in season 4 (ranging from about 6 to 30 days), with masses following the same trend (ranging from \textrm{$5.0 \times 10^{6} M_\odot$} to \textrm{$2.2 \times 10^{7} M_\odot$}). Though the lags change by quite a bit, there is not a significant luminosity change between seasons. Season 1 is our preferred mass. Seasons 1 \& 4 show blue asymmetry while season 3 shows red asymmetry. 
The velocity-resolved time lags for season 3 are suggestive of a disk and are indeterminate for seasons 1 \& 4 (see Figure~\ref{figure:v_resolved3} and Figure~\ref{figure:v_resolved4}).

\subsection{PG 2304+042} \label{subsec:pg}
See the light curve, CCFs, CCCDs, mean, and rms spectra for this object in Figure~\ref{figure:pg2304_lc}. The only Palomar-Green quasar in this paper, the H$\beta$ profile is asymmetric with a bump on the red wing.
We have three seasons of data on this object, starting in 2020. All three seasons display sufficient variability in order to measure a time delay and the delays agree for each season. The measured masses also agree well. The kinematics from the velocity-resolved results suggest a disk in season 1, are flat in season 2, and suggest outflow in season 3 (see Figure~\ref{figure:v_resolved4}). All three seasons show red asymmetry with season 1 having the strongest asymmetry.

\subsection{NGC 7603} \label{subsec:ngc7603}
See the light curve, CCFs, CCCDs, mean, and rms spectra for this object in Figure~\ref{figure:ngc7603_lc}. This object, also known as Mrk 530, has a history of variable and asymmetric H$\beta$ profiles \citep{Goodrich1989, Kollatschny2000}. This object has moderate Fe II emission in its optical spectrum. See Section~\ref{subsec:lightcurves} for an explanation of how we addressed the possible Fe II emission contaminant when constructing our H$\beta$ light curves.
This object has one season of data. The velocity-resolved results indicate a disk-like structure (see Figure~\ref{figure:v_resolved4}). This season shows red asymmetry.

\section{Conclusions} \label{sec:con}
In this fourth paper in the MAHA series we report RM results, including velocity-resolved time lags, for 14 objects without prior RM results. We present multiple seasons of data for 9 of the 14 objects, allowing us to investigate time lag and BLR kinematic evolution over a period of several years. We find evidence for changing time lags without corresponding luminosity changes as well as changing line asymmetries and BLR kinematics as revealed by velocity-resolved lags. We observe predominantly red asymmetries. We observe mostly disk-like BLR kinematics, with 12 out of 33 observed seasons displaying disk-like kinematics. Five out of the 33 seasons show outflow, two show inflow, and two show the same time lag across the line profile. 12 out of the 33 have kinematics that are not determinable from their velocity-resolved time lags. 

\begin{acknowledgments}
We thank the referee for their valuable suggestions. We thank WIRO engineers James Weger, Conrad Vogel, and Andrew Hudson for their indispensable and invaluable assistance. LCH was supported by the National Science Foundation of China (11721303, 11991052, 12011540375, 12233001), the National Key R\&D Program of China (2022YFF0503401), and the China Manned Space Project (CMS-CSST-2021-A04, CMS-CSST-2021-A06). This work is supported by the National Science Foundation under REU grant AST 1852289. PD acknowledges financial support from NSFC through grants NSFC-12022301 and -11873048, and from National Key R\&D Program of China grant 2021YFA1600404. TEZ acknowledges support from NSF grant 1005444I. This research has made use of the NASA/IPAC Extragalactic Database (NED), which is funded by the National Aeronautics and Space Administration and operated by the California Institute of Technology. 
This research has made use of the SIMBAD database, operated at CDS, Strasbourg, France. The ztfquery code was funded by the European Research Council (ERC) under the European Union's Horizon 2020 research and innovation programme (grant agreement n°759194 - USNAC, PI: Rigault). This work is also based on observations obtained with the Samuel Oschin 48 inch Telescope at the Palomar Observatory as part of the Zwicky Transient Facility project. ZTF is supported by the National Science Foundation under grant No. AST-1440341 and a collaboration including Caltech, IPAC, the Weizmann Institute for Science, the Oskar Klein Center at Stockholm University, the University of Maryland, the University of Washington, Deutsches Elektronen-Synchrotron and Humboldt University, Los Alamos National Laboratories, the TANGO Consortium of Taiwan, the University of Wisconsin at Milwaukee, and Lawrence Berkeley National Laboratories. Operations are conducted by COO, IPAC, and UW. 
\end{acknowledgments}

\software{Astropy \citep{astropy:2013, astropy:2018, astropy:2022},
Scipy \citep{2020SciPy},
Numpy \citep{Numpy},
Matplotlib \citep{Matplotlib},
LMFIT (\citep{lmfit}),
Pandas (\url{https://github.com/pandas-dev/pandas}),
dust\_extinction (\href{https://github.com/karllark/dust_extinction}{https://github.com/karllark/dust\_extinction}),
ztfquery \citep{Rigault2018},
PyCALI \citep{Li2014}
}

\bibliography{references}{}
\bibliographystyle{aasjournal}

\end{document}